\documentclass{research4cacm}

\usepackage{times}
\usepackage{helvet}
\usepackage{amsmath} 
\usepackage{amssymb}
\usepackage{amsfonts}
\usepackage{algorithm}
\usepackage{algpseudocode}
\usepackage{subcaption}
\usepackage{graphicx}
\usepackage{epstopdf}
\usepackage{booktabs} 
\usepackage{amsopn}
\usepackage{comment}
\usepackage{enumitem}
\usepackage{makebox}
\usepackage{url}
\usepackage{color}
\usepackage{multirow}
\usepackage{booktabs}
\usepackage{array}
\usepackage{fixltx2e}
\usepackage{datetime}

\newtheorem{challenge}{Challenge}
\newtheorem{property}{Property}
\newtheorem{definition}{Definition}

\newcommand{\argmax}{\mathop{\mathrm{arg\,max}}}
\begin{document}
%

\title{The Use of Bandit Algorithms in Intelligent Interactive Recommender Systems
%
}
%
%
%
%
%
\numberofauthors{2} 
%
\author{
%
%
\alignauthor
Qing Wang\\
       \affaddr{IBM Research AI}\\
       \email{qing.wang1@ibm.com}
}

\maketitle

\begin{abstract}
In today's business marketplace, many high-tech Internet enterprises constantly explore innovative ways to provide optimal online user experiences for gaining competitive advantages. The great needs of developing intelligent interactive recommendation systems are indicated, which could sequentially suggest users the most proper items by accurately predicting their preferences, while receiving the up-to-date feedback to refine the recommendation results, continuosly. Multi-armed bandit algorithms, which have been widely applied into various online systems, are quite capable of delivering such efficient recommendation services. However, few existing bandit models are able to adapt to new changes that introduced by modern recommender systems.
\end{abstract}

\section{Introduction} \label{sec:introduction}
Facing the fast growing development of online services, recommender systems have been widely explored and increasingly become a popular research area in recent decades. In order to boost sales as well as improve users' visiting experience, many practical applications in major companies (e.g. Google, Amazon, Netflix, and etc.) provide efficient online recommendation services to help consumers deal with the overwhelming information. Most recently, interactive recommender systems have emerged striving to promptly feed an individual with proper items (e.g., news articles, music, movies, and etc.) according to the current context, adaptively optimize the underlying recommendation model using the up-to-date feedback and continuously maximize his/her satisfaction in a long run~\cite{zhao2013interactive}. To achieve this goal, it becomes a critical task for modern recommender systems to identify the goodness of match between users' preferences and target items.

However, identifying an appropriate match between user preferences and target items is quite difficult, especially with the well-known cold-start problem. Since a significant number of users/items might be completely new to the system with no consumption history at all, the cold-start problem~\cite{schein2002methods} makes recommender systems ineffective unless they collects more additional information~\cite{chang2015space}. Generally, the cold-start issue is often referred to as an exploration / exploitation dilemma: maximizing user satisfaction based on their consumption history, while gathering new information for improving the goodness of match between user preferences and items~\cite{li2010contextual}. Besides, successful recommender systems are required to adaptively predict a user's preference by making use of the user's up-to-date feedback (e.g., likes or dislikes) on recommended items as well as the observed context. This can be naturally modeled as contextual bandit problems (e.g., LinUCB~\cite{li2010contextual} and Thompson sampling~\cite{chapelle2011empirical}), where each arm corresponds to an item, pulling an item indicates recommending an item, and the reward is the instant feedback from a user after the recommendation. Contextual bandit algorithms have been widely applied in various interactive recommender systems by achieving an optimal tradeoff between \textit{exploration} and \textit{exploitation}.
Based on the preliminary studies~\cite{jannach2016recommender,li2010contextual,adomavicius2005toward}, several practical challenges are identified in modern recommender systems.

\begin{challenge}\label{challenge2}
\textit{How do we promptly capture both of the varying popularity of item content and the evolving customer preferences over time, and further utilize them for recommendation improvement?}
\end{challenge}
Existing contextual bandit algorithms take the observed contextual information as the input and predict the expected reward for each arm with an assumption that the reward is invariant under the same context. However, this assumption rarely holds in practice since the real-world problems often involve some underlying processes that dynamically evolving overtime~\cite{zheng2018drn}. For example, the popularity of an item (e.g., a news article or movie) usually drops down quickly after its first publication, while user interests may evolve after exploring new emerged items (e.g., music or video games). Since both of the popularity of item content and user preferences are dynamically evolving over time, a new challenge is introduced requiring the system instantly tracks these changes, i.e., the time varying behaviors of the reward. To overcome this challenge, we propose a dynamical context drift model based on particle learning~\cite{zeng2016online}, where the dynamic behaviors of the reward is explicitly modeled as a set of random walk particles.
\begin{challenge}\label{challenge3}
 \textit{How to do we effectively model the dependencies among items and incorporate them into bandit algorithms?}
\end{challenge}
In multi-armed bandit problems, many policies~\cite{LangfordZ07,chapelle2011empirical,auer2002using} have been come up with by assuming that the success probability of each arm are independent~\cite{pandey2007multi}. However, in practical recommender systems, a user's implicit feedback (e.g., rating or click)~\cite{hu2008collaborative} on one recommended item may infer the user's preference on the other items. For example, news articles with similar topics (e.g., sports, politics, etc.) and publisher information are likely to receive the similar feedback based on a user's preference. In the other words, the dependencies among arms can be utilized for reward prediction improvement and further facilitate the maximization the users' satisfaction in a long run. Therefore, we propose hierarchical multi-armed bandit algorithms to exploit the dependencies among arms organized in the form of the hierarchies.

The remainder of this paper is organized as follows. In Section~\ref{sec:relatedwork}, we provide a brief summary of the state-of-the-art literature. The mathematical formalizations of the aforementioned problems are given in Section~\ref{sec:formulation}. The solutions and methodologies are presented in Section~\ref{sec:solution}. In Section~\ref{sec:experiment}, we show the experimental results. Finally, we will discuss the future work in Section~\ref{sec:conclusion}.

\section{Related Work} \label{sec:relatedwork}
In this section, we highlight these existing work related to our approaches in this section.

\subsection{Contextual Multi-armed Bandit}
Interactive recommender systems play an essential role in our daily life due to the abundance of online services~\cite{zhao2013interactive} in this information age. For an individual, the systems can continuously refine the recommended results by making use of the up-to-date feedback according to the current context including both user and item content information. The cold-start problem inherent in learning from interactive feedback has been well dealt with using contextual multi-armed bandit algorithms~\cite{LangfordZ07,li2010contextual,chapelle2011empirical,tang2015personalized,hill2017efficient}, which have been widely applied into personalized recommendation services. In~\cite{li2010contextual}, LinUCB is proposed to do personalized recommendation on news article. Tang et al.~\cite{tang2015personalized} come up with a novel parameter-free algorithm based on a principled sampling approach. In~\cite{hill2017efficient}, authors present an efficient contextual bandit algorithm for realtime multivariate optimization on large decision spaces. Different from these existing algorithms assuming that the reward is invariant under the same context, a context drift model is proposed to deal with the contextual bandit problem by considering the dynamic behaviors of reward into account.

Besides, these prior work assumes arms are independent, which neither holds true in reality. Since the real-world items tend to be correlated with each other, a delicate framework~\cite{pandey2007multi} is developed to study the bandit problem with dependent arms. In light of the topic modeling techniques, Wang et al.~\cite{wang2017online} come up with a new generative model to explicitly formulate the item dependencies as the clusters on arms. Pandey et al.~\cite{pandey2007bandits} uses the taxonomy structure to exploit dependencies among the arm in the context-free bandit setting. CoFineUCB approach in~\cite{yue2012hierarchical} is proposed to utilize a coarse-to-fine feature hierarchy to reduce the cost of exploration, where the hierarchy was estimated by a small number of existing user profiles.  Some other recent studies explore the bandit dependencies for a group recommendation delivery by assuming users in the same group react with similar feedback to the same recommended item~\cite{gentile2014online,wu2016contextual,wang2017factorization,wang2017interactive,zhou2016latent}. In contrast, we study the contextual bandit problem with a given hierarchical structure of arms, where the hierarchy is constructed based on the category of each item.

\subsection{Sequential Online Inference}
Sequential online inference has been applied to infer the latent states and learn unknown parameters of our context drift model. Sequential monte carlo sampling~\cite{halton1962sequential} and particle learning~\cite{carvalho2010particle} are two popular sequential learning methods~\cite{zeng2016onlinebigdata}.
Sequential Monte Carlo (SMC) methods consist of a set of Monte Carlo methodologies to solve the filtering problem~\cite{doucet2000sequential}, which provides a set of simulation-based methods for computing the posterior distribution. These methods allow inference of full posterior distributions in general state space models, which may be both nonlinear and non-Gaussian. Particle learning provides state filtering, sequential parameter learning and smoothing in a general class of state space models~\cite{carvalho2010particle}. Particle learning is for approximating the sequence of filtering and smoothing distributions in light of parameter uncertainty for a wide class of state space models. The central idea behind particle learning is the creation of a particle algorithm that directly samples from the particle approximation to the joint posterior distribution of states and conditional sufficient statistics for fixed parameters in a fully-adapted \texttt{resample}-\texttt{propagate} framework.

\section{Problem Formulation}\label{sec:formulation}
In this section, we first formally define the contextual multi-armed bandit problem, and then provide the dynamic context drift modeling and hierarchical dependency modeling. Some important notations mentioned in this paper are summarized in Table~\ref{tab:notations}.

\begin{table}[!h]
	\vskip -0.1in
    \small
	\centering
	\caption{Important Notations}
	\vskip -0.1in
	\begin{tabular}{lp{6.5cm}}
		\toprule
		\textbf{Notation} & \textbf{Description}\\
		\midrule
		$a^{(i)}$       & the $i$-th arm.  \\
		$\mathcal{A}$   & the set of arms, $\mathcal{A} =  \{a^{(1)},...,a^{(K)}\}$. \\
        $\mathcal{H}$   & the constructed hierarchy. \\
        $\mathcal{X}$   & the $d$-dimensional context feature space. \\
		$\mathbf{x}_t$  & the context at time $t$, and represented by a vector. \\
		$r_{k,t}$     & the reward of pulling the arm $a^{(k)}$ at time $t$, $a^{(k)} \in \mathcal{A}$. \\
		$y_{k,t}$  & the predicted reward for the arm $a^{(k)}$ at time $t$. \\
        $\mathcal{P}_k$ & the set of particles for the arm $a^{(k)}$ and $\mathcal{P}^{(i)}_k$ is the $i^{th}$ particle of $\mathcal{P}_k$. \\
        $\mathbf{S}_{\pi,t}$ & the sequence of $(\mathbf{x}_i,\pi(\mathbf{x}_i),r_{\pi(\mathbf{x}_i)})$ observed until time $t$. \\
        $\mathbf{w}_k$ & the coefficient vector used to predict reward of the arm $a^{(k)}$. \\
        $\mathbf{c}_{\mathbf{w}_k}$ & the constant part of $\mathbf{w}_k$. \\
        $\delta_{\mathbf{w},t}$ & the drifting part of $\mathbf{w}_k$ at time $t$. \\
        $\eta_{k,t}$ & the standard Gaussian random walk at time $t$, given $\eta_{k,t-1}$.\\
        $\theta_k$ & the scale parameters used to compute $\delta_{\mathbf{w},t}.$ \\
        $\pi$ & the policy for pulling arm sequentially. \\
        $R_\pi$ & the cumulative reward of the policy $\pi$. \\
        $f_{a^{(k)}}(\mathbf{x}_t)$ & the reward prediction function of the arm $a^{(k)}$, given context $\mathbf{x}_t$. \\
        $\sigma^2_k$ & the variance of reward prediction for the arm $a^{(k)}$. \\
        $\alpha$, $\beta$ & the hyper parameters determine the distribution of $\sigma^2_k$. \\
        $\mathbf{\mu_w}$,$\mathbf{\Sigma_w}$ & the hyper parameters determine the distribution of $\mathbf{w}_k$. \\
        $\mathbf{\mu_c}$, $\mathbf{\Sigma_c}$ & the hyper parameters determine the distribution of $\mathbf{c}_{\mathbf{w}_k}$. \\
        $\mathbf{\mu_\theta}$, $\mathbf{\Sigma_\theta}$ & the hyper parameters determine the distribution of $\mathbf{\theta}_k$. \\
        $\mathbf{\mu_\eta}$, $\mathbf{\Sigma_\eta}$ & the hyper parameters determine the distribution of $\mathbf{\eta}_{k,t}$. \\
		\bottomrule
	\end{tabular}
	\label{tab:notations}
\end{table}

\subsection{Basic Concepts and Terminologies}\label{sec:base}

Let $\mathcal{A} = \{a^{(1)},a^{(2)} ..., a^{(K)}\}$ be a set of arms, where $K$ is the number of arms. A $d$-dimensional feature vector $\mathbf{x}_t\in \mathcal{X}$ represents the contextual information at time $t$, and $\mathcal{X}$ is the $d$-dimensional feature space. Generally, the contextual multi-armed problem involves a series of decisions over a finite but possibly unknown time horizon $T$. A policy $\pi$ makes a decision at each time $t=[1,\dots,T]$ to select the arm $\pi(\mathbf{x}_t) \in \mathcal{A}$ to pull based on the contextual information $\mathbf{x}_t$. After pulling an arm, the policy receives a reward from the selected arm. The reward of an arm $a^{(k)}$ at time $t$ is denoted as $r_{k,t}$, whose value is drawn from an unknown distribution determined by the context $x_t$ presented to arm $a^{(k)}$. However the reward $r_{k,t}$ is unobserved unless arm $a^{(k)}$ is pulled. The total reward received by the policy $\pi$ is
\begin{equation}
\small
R_\pi=\sum_{t=1}^T{r_{\pi(\mathbf{x}_t)}}.
\end{equation}

The goal is to identify the optimal policy $\pi^*$ for maximizing the total reward after $T$ iterations.
\begin{equation}
  \small
  \pi^* = \argmax_{\pi}E(R_{\pi}) = \argmax_{\pi} \sum_{t=1}^{T} E(r_{\pi(\mathbf{x}_t)}|t).
  \label{eq:objective}
\end{equation}

Before selecting one arm at time $t$, a policy $\pi$ typically learns a model to predict the reward for each arm according to the historical observation, $S_{\pi,t-1}=\{(\mathbf{x}_i,\pi(\mathbf{x}_i),r_{\pi(\mathbf{x}_i)})|1\le i < t\}$, which consists of a sequence of triplets.
The reward prediction helps the policy $\pi$ make decisions to increase the total reward.

 Assume $y_{k,t}$ is the predicted reward of arm $a^{(k)}$, which is determined by
 \begin{equation}
 \small
   y_{k,t} = f_{a^{(k)}}(\mathbf{x}_t),
 \end{equation}
 where the context $\mathbf{x}_t$ is the input and $f_{a^{(k)}}$ is the reward mapping function for arm $a^{(k)}$.
\begin{figure*}[htp!]
    \begin{subfigure}{0.5\textwidth}
    \centering
    \scalebox{0.45}{
        \includegraphics{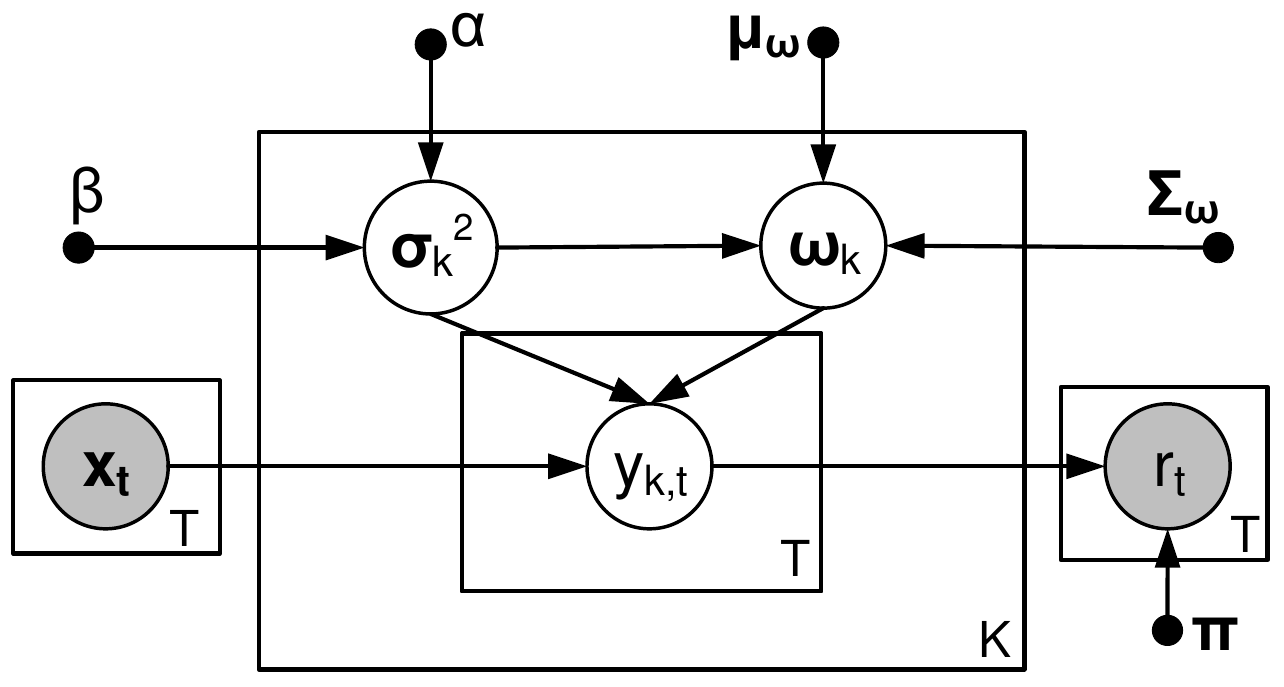}
    }
   \caption{Multi-armed bandit problem.~\cite{zeng2016online}}
   \label{fig:modelold}
    \end{subfigure}
    \begin{subfigure}{0.5\textwidth}
        \centering
        \scalebox{0.4}{
        \includegraphics{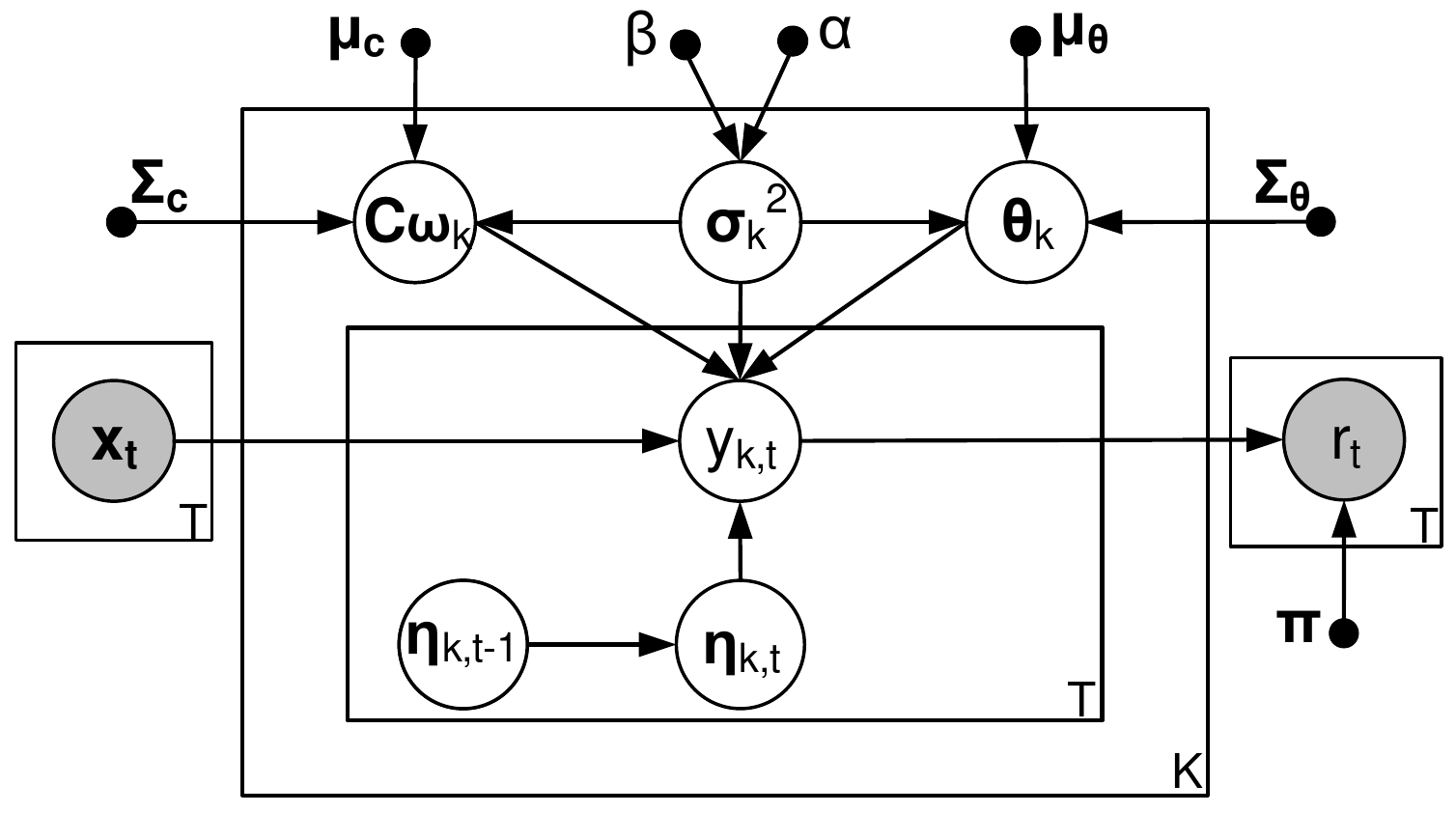}
        }
        \caption{Time varying multi-armed bandit problem.~\cite{zeng2016online}}
        \label{fig:modelnew}
    \end{subfigure}
    \vspace{-0.1in}
    \caption{Graphical model representation for bandit problems. Random variable is denoted as a circle. The circle with gray color filled means the corresponding random variable is observed. Black dot represents a hyper parameter.}
    \vspace{-0.1in}
\end{figure*}
One popular mapping function is defined as the linear combination of the feature vector $\mathbf{x}_t$, which has been successfully used in bandit problems~\cite{li2010contextual}\cite{agrawal2013thompson}.
Specifically, $f_{a^{(k)}}(\mathbf{x}_t)$ is formally given as follows:
\begin{equation}
    \small
    \label{eq:regression}
  f_{a^{(k)}}(\mathbf{x}_t) = \mathbf{x}^\intercal_t\mathbf{w}_k + \varepsilon_k,
\end{equation}
where $\mathbf{x}^\intercal_t$ is the transpose of contextual information $\mathbf{x}_t$, $\mathbf{w}_k$ is a $d$-dimensional coefficient vector, and $\varepsilon_k$ is a zero-mean Gaussian noise with variance $\sigma^2_k$, i.e., $\varepsilon_k \thicksim \mathcal{N}(0,\sigma^2_k)$. 
Accordingly,
\begin{equation}
    \small
    \label{eq:genY}
    y_{k,t} \thicksim \mathcal{N}(\mathbf{x}_t^\intercal\mathbf{w}_k,\sigma^2_k).
\end{equation}
In this setting, a graphical model representation is provided in Figure~\ref{fig:modelold}.
The context $\mathbf{x}_t$ is observed at time $t$. The predicted reward value $y_{k,t}$ depends on random variable $\mathbf{x}_t$, $\mathbf{w}_k$, and $\sigma^2_k$.
 A conjugate prior distribution for the random variables $\mathbf{w}_k$ and $\sigma^2_k$ is assumed and defined as \underline{N}ormal \underline{I}nverse \underline{G}amma ($\mathcal{NIG}$) distribution with the hyper parameters $\mathbf{\mu_{w}}$, $\mathbf{\Sigma_{w}}$, $\alpha$, and $\beta$. The distribution is denoted as $\mathcal{NIG}(\mathbf{\mu_{w}},\mathbf{\Sigma_{w}},\alpha,\beta)$ and shown below:
\begin{equation}
    \small
    \begin{split}
  \mathbf{w}_k|\sigma^2_k \thicksim \mathcal{N}(\mathbf{\mu_{w}},\sigma^2_k\mathbf{\Sigma_{w}}), \\
  \sigma^2_k \thicksim \mathcal{IG}(\alpha,\beta),
  \end{split}
  \label{eq:nig}
\end{equation}
where the hyper parameters are predefined.

 A policy $\pi$ selects one arm $a^{(k)}$ to pull according to the reward prediction model. After pulling arm $a^{(k)}$ at time $t$, a corresponding reward $r_{k,t}$ is observed, while the rewards of other arms are still hidden. A new triplet $(\mathbf{x}_t,\pi(\mathbf{x}_t),r_{\pi(\mathbf{x}_t)})$ is obtained and a new sequence $S_{\pi,t}$ is formed by combining $S_{\pi,t-1}$ with the new triplet. The posterior distribution of $\mathbf{w}_k$ and $\sigma^2_k$ given $S_{\pi,t}$  is a $\mathcal{NIG}$ distribution. Denoting the parameters of $\mathcal{NIG}$ distribution at time $t-1$ as $\mathbf{\mu_{w_{t-1}}}$, $\mathbf{\Sigma_{w_{t-1}}}$, $\alpha_{t-1}$, and $\beta_{t-1}$, the hyper parameters at time $t$ are updated as follows:
 \begin{equation}
 \small
 \begin{split}
   &\mathbf{\Sigma_{w_{t}}} = (\mathbf{\Sigma_{w_{t-1}}}^{-1} + \mathbf{x}_t\mathbf{x}_t^\intercal)^{-1}, \\
   &\mathbf{\mu_{w_{t}}} = \mathbf{\Sigma_{w_{t}}}(\mathbf{\Sigma_{w_{t-1}}}^{-1}\mathbf{\mu_{w_{t-1}}} + \mathbf{x}_tr_{\pi(\mathbf{x}_t)}),
   \ \ \ \alpha_t = \alpha_{t-1} + \frac{1}{2},\\
   &\beta_t = \beta_{t-1} + \frac{1}{2}[r^2_{\pi(\mathbf{x}_t)}+\mathbf{\mu^\intercal_{w_{t-1}}}\mathbf{\Sigma_{w_{t-1}}}^{-1}\mathbf{\mu_{w_{t-1}}} - \mathbf{\mu^\intercal_{w_t}}\mathbf{\Sigma_{w_t}}^{-1}\mathbf{\mu_{w_t}}].
   \end{split}
\label{eq:posteiorUpdate}
 \end{equation}

Note that, the posterior distribution of $\mathbf{w}_k$ and $\sigma^2_k$ at time $t-1$ is considered as the prior distribution at time $t$.
Both LinUCB~\cite{li2010contextual} and Thompson Sampling~\cite{chapelle2011empirical} will be incorporated into our dynamic context drift model to address the contextual multi-armed bandit problem. More details will be discussed in Section 4 after modeling the context drift.

\subsection{Dynamic Context Drift Modeling}
As mentioned in Section 3.1, the reward prediction for arm $a^{(k)}$ is estimated by a linear combination of contextual features $\mathbf{x}_t$, with coefficient vector $\mathbf{w}_k$. Each element in the coefficient vector $\mathbf{w}_k$ indicates the contribution of the corresponding feature 
for reward prediction.
The aforementioned model is based on the assumption that $\mathbf{w}_k$ is unknown but fixed~\cite{agrawal2013thompson}, which rarely holds in practice. The real-world problems often involve some underlying processes. These processes often lead to the dynamics in the contribution of each context feature to the reward prediction. To account for the dynamics,
our goal is to come up with a model having the capability of capturing the drift of $\mathbf{w}_k$ over time and subsequently obtain a better fitted model for the dynamic reward change.
Let $\mathbf{w}_{k,t}$ denote the coefficient vector for arm $a^{(k)}$ at time $t$. Taking the drift of $\mathbf{w}_k$ into account,  $\mathbf{w}_{k,t}$ is formulated as follows:
\begin{equation}
    \label{eq:wdecompose}
  \mathbf{w}_{k,t} = \mathbf{c}_{\mathbf{w}_{k}} + \delta_{\mathbf{w}_{k,t}},
\end{equation}
where $\mathbf{w}_{k,t}$ is decomposed into two components including both the stationary component $\mathbf{c}_{\mathbf{w}_{k}}$ and the drift component $\delta_{\mathbf{w}_{k,t}}$. Both components are $d$-dimensional vectors. Similar to modeling $\mathbf{w}_k$ in Figure~\ref{fig:modelold}, the stationary component $\mathbf{c}_{\mathbf{w}_{k}}$ can be generated with a conjugate prior distribution
 \begin{equation}
  \mathbf{c}_{\mathbf{w}_{k}} \thicksim \mathcal{N}(\mathbf{\mu_c},\sigma^2_k\mathbf{\Sigma_c}),
\end{equation}
where $\mathbf{\mu_c}$ and $\mathbf{\Sigma_c}$ are predefined hyper parameters as shown in Figure~\ref{fig:modelnew}.

However, it is difficult to model the drift component $\delta_{\mathbf{w}_{k,t}}$  with a single function due to the diverse characteristics of the context. For instance, given the same context, the CTRs of some articles change quickly, while some articles may have relatively stable CTRs. Moreover, the coefficients for different elements in the context feature can change with diverse scales.  To simplify the inference, we assume that each element of $\delta_{\mathbf{w}_{k,t}}$  drifts independently. Due to the uncertainty of drifting, we formulate $\delta_{\mathbf{w}_{k,t}}$ with a standard Gaussian random walk $\mathbf{\eta_{k,t}}$ and a scale variable $\mathbf{\theta}_k$ using the following Equation: 
\begin{equation}
\label{eq:drift}
 \delta_{\mathbf{w}_{k,t}} =  \mathbf{\theta}_k\odot\mathbf{\eta_{k,t}},
\end{equation}
where $\mathbf{\eta_{k,t}}\in\mathcal{R}^d$ is the drift value at time $t$ caused by the standard random walk and $\theta_k\in\mathcal{R}^d$ contains the changing scales for all the elements of $\delta_{\mathbf{w}_{k,t}}$. The operator $\odot$ is used to denote the element-wise product. The standard Gaussian random walk is defined with a Markov process as shown in Equation~\ref{eq:randomwalk}.
\begin{equation}
\label{eq:randomwalk}
  \mathbf{\eta_{k,t}} = \mathbf{\eta_{k,t-1}} + \mathbf{v},
\end{equation}
where $\mathbf{v}$ is a standard Gaussian random variable defined by $\mathbf{v} \thicksim \mathcal{N}(\mathbf{0},\mathbf{I}_d)$, and $\mathbf{I}_d$ is a $d\times d$-dimensional identity matrix. It is equivalent that $\mathbf{\eta_{k,t}}$ is drawn from the Gaussian distribution
\begin{equation}
  \mathbf{\eta_{k,t}} \thicksim \mathcal{N}(\mathbf{\eta_{k,t-1}},\mathcal{I}_d).
\end{equation}

The scale random variable $\theta_k$ is generated with a conjugate prior distribution
\begin{equation}
  \theta_k \thicksim \mathcal{N}(\mathbf{\mu_\theta}, \sigma^2_k\mathbf{\Sigma_\theta}),
\end{equation}
where $\mu_\theta$ and $\mathbf{\Sigma_\theta}$ are predefined hyper parameters. $\sigma^2_k$ is drawn from the Inverse Gamma (abbr., $\mathcal{IG}$) distribution provided in Equation~\ref{eq:nig}.
Combining Equations \ref{eq:wdecompose} and \ref{eq:drift}, we obtain
\begin{equation}
  \mathbf{w}_{k,t} = \mathbf{c}_\mathbf{w_{k}} + \mathbf{\theta}_k\odot\mathbf{\eta_{k,t}}.
   \label{eq:weights}
\end{equation}

According to Equation~\ref{eq:regression}, $y_{k,t}$ can be computed as
\begin{equation}
  y_{k,t} = \mathbf{x}_t^\intercal(\mathbf{c}_\mathbf{w_k} + \mathbf{\theta}_k\odot\mathbf{\eta_{k,t}}) + \epsilon_k.
\end{equation}

Accordingly, $y_{k,t}$ is modeled to be drawn from the following Gaussian distribution:
\begin{equation}
    \label{eg:genYNew}
  y_{k,t} \thicksim \mathcal{N}(\mathbf{x}_t^\intercal(\mathbf{c}_\mathbf{w_k} + \mathbf{\theta}_k\odot\mathbf{\eta_{k,t}}),\sigma^2_k).
\end{equation}

The new context drift model is presented with a graphical model representation in Figure~\ref{fig:modelnew}. Compared with the model in Figure~\ref{fig:modelold}, a standard Gaussian random walk $\eta_{k,t}$ and the corresponding scale $\theta_k$ for each arm $a^{(k)}$ are introduced in the new model. The new model explicitly formulates the drift of the coefficients for the reward prediction, considering the dynamic behaviors of the reward in real-world application. From the new model, each element value of $\mathbf{c}_\mathbf{w_{k}}$ indicates the contribution of its corresponding feature in predicting the reward, while the element values of $\mathbf{\theta}_k$ show the scales of context drifting for the reward prediction. A large element value of $\mathbf{\theta}_k$ signifies a great context drifting occurring to the corresponding feature over time.

\subsection{Hierarchical Dependency Modeling}
Generally, the items (e.g., news articles, ads, etc.) in real recommendation services (e.g., news recommendation, ads recommendation, etc.) can be categorized using a predefined taxonomy. By encoding these prior knowledge, it allows us to reformulate the contextual bandit problem with dependent arms organized hierarchically, which explores the arm's feature space from a coarse to fine level.

Let $\mathcal{H}$ denote the taxonomy, which contains a set of nodes (i.e., items) organized in a tree-structured hierarchy.
Given a node $a^{(i)}\in\mathcal{H}$, $pa(a^{(i)})$ and $ch(a^{(i)})$ represent the parent and children sets, respectively.
Accordingly, Property~\ref{prop:hierarchy} is given as follows:
 \begin{property}
   If $pa(a^{(i)})=\emptyset$, then node $a^{(i)}$ is the root node. 
   If $ch(a^{(i)})=\emptyset$, then $a^{(i)}$ is a leaf node, which represents an item.
   Otherwise, $a^{(i)}$ is a category node when $ch(a^{(i)})\neq\emptyset.$
   \label{prop:hierarchy}
 \end{property}
Since the goal is to recommend a proper item for an individual
and only a leaf node of $\mathcal{H}$ represents an item,
the recommendation process cannot be completed until a leaf node is selected at each time $t=[1,\dots,T]$.
Therefore, the contextual bandit problem with dependent arms is reduced to the optimal selection of a path in $\mathcal{H}$ from the root to a leaf node,
and multiple arms along the path are sequentially selected based on
the contextual vector $\mathbf{x}_t$ at time $t$.

Let $pth(a^{(i)})$ be a set of nodes along the path
from the root node to the leaf node $a^{(i)}$ in $\mathcal{H}$.
Assume that $\pi_\mathcal{H}(\mathbf{x}_t|t)$ is the path selected by
the policy $\pi$ in light of the contextual information $\mathbf{x}_t$ at time $t$.
For every arm selection policy $\pi$ we have:
\begin{property}
Given the contextual information $\mathbf{x}_t$ at time $t$,
if a policy $\pi$ selects a node $a^{(i)}$ in the hierarchy $\mathcal{H}$
and receives positive feedback (i.e., success), the policy $\pi$
receives positive feedback as well by selecting the nodes along the path $pth(a^{(i)})$.
\label{prop:hierarchyconstraint}
\end{property}

Let $r_{\mathbf{x}_t, \pi_\mathcal{H}(\mathbf{x}_t|t)}$ denote the reward obtained
by the policy $\pi$ after selecting multiple arms along the path
$\pi_\mathcal{H}(\mathbf{x}_t|t)$ at time $t$.
The reward is computed as follows:
\begin{equation}
  \small
  r_{\mathbf{x}_t, \pi_\mathcal{H}(\mathbf{x}_t|t)}=
  \sum_{a^{(i)}\in\pi_\mathcal{H}(\mathbf{x}_t|t),ch(a^{(i)})\neq\emptyset}{r_{\mathbf{x}_t, \pi(\mathbf{x}_t|ch(a^{(i)}))}},
  \label{eq:one-time-reward}
\end{equation}
where $\pi(\mathbf{x}_t|ch(a^{(i)}))$ represents the arm selected from the children of
 $a^{(i)}$, given the contextual information $\mathbf{x}_t.$
After $T$ iterations, the total reward received by the policy $\pi$ is:
\begin{equation}
\small
R_{\pi_{\mathcal{H}}} = \sum_{t=1}^{T}{r_{\mathbf{x}_t, \pi_\mathcal{H}(\mathbf{x}_t|t)}}.
  \label{eq:total-reward}
\end{equation}
The optimal policy $\pi^*$ with respect to $\mathcal{H}$ is determined by
\begin{equation}
\small
  \pi^* = \argmax_{\pi}E(R_{\pi_\mathcal{H}}) = \argmax_{\pi} \sum_{t=1}^{T} E(r_{\mathbf{x}_t, \pi_\mathcal{H}(\mathbf{x}_t)}|t).
\end{equation}
The reward prediction for each arm is conducted by Equation~\ref{eq:rewardPrediction1}, and then the optimal policy can be equivalently determined by
\begin{equation}
  \pi^* =
   \argmax_{\pi} \sum_{t=1}^{T}{\sum_{ \substack{a^{(i)}\in\pi_\mathcal{H}(\mathbf{x}_t|t), \\ ch(a^{(i)})\neq\emptyset}}{E_{\theta_{\pi(\mathbf{x}_t|ch(a^{(i)}))}}}(\mathbf{x}_t^T\theta_{\pi(\mathbf{x}_t|ch(a^{(i)}))}|t)}.
\end{equation}

Three popular bandit algorithms (i.e., $\epsilon$-greedy~\cite{tokic2010adaptive}, Thompson sampling~\cite{chapelle2011empirical}, and LinUCB~\cite{li2010contextual}) are incorporated with our proposed model. Since our model explicitly makes use of the prior knowledge (i.e., hierarchies), which allows it to converge much faster by exploring the item's feature space hierarchically.

\section{Solution and Algorithm}\label{sec:solution}
In this section, we will provide the solutions and algorithms for the aforementioned modeling problems, respectively.
\subsection{Solution to Context Drift Modeling}
In this section, we present the methodology for online inference of the context drift model.

The posterior distribution inference involves four random variables, i.e., $\sigma^2_k$, $\mathbf{c}_\mathbf{w_{k}}$, $\mathbf{\theta}_k$, and $\mathbf{\eta_{k,t}}$. According to the graphical model in Figure~\ref{fig:modelnew}, the four random variables are grouped into two categories: parameter random variable and latent state random variable. $\sigma^2_k$, $\mathbf{c}_\mathbf{w_{k}}$, $\mathbf{\theta}_k$ are parameter random variables since they are assumed to be fixed but unknown, and their values do not depend on the time. Instead, $\mathbf{\eta_{k,t}}$ is referred to as a latent state random variable since it is not observable and its value is time dependent according to Equation~\ref{eq:randomwalk}. After pulling the arm $a^{(k)}$ according to the context $\mathbf{x}_t$ at time $t$, a reward is observed as $r_{k,t}$. Thus, $\mathbf{x}_t$ and $r_{k,t}$ are referred to as observed random variables.
Our goal is to infer both latent parameter variables and latent state random variables to sequentially fit the observed data. However, since the inference partially depends on the random walk which generates the latent state variable, we use the sequential sampling based inference strategy that are widely used sequential monte carlo sampling~\cite{smith2013sequential}, particle filtering~\cite{djuric2003particle}, and particle learning~\cite{carvalho2010particle} to learn the distribution of both parameter and state random variables.

Since state $\mathbf{\eta_{k,t-1}}$ changes over time with a standard Gaussian random walk, it follows a Gaussian distribution after accumulating $t-1$ standard Gaussian random walks. Assume $\mathbf{\eta_{k,t-1}} \thicksim \mathcal{N}(\mu_{\eta_{k}},\Sigma_{\eta_{k}})$, a particle is defined as follows.

\begin{definition}[Particle] A particle of an arm $a^{(k)}$ is a container which maintains the current status information of $a^{(k)}$. The status information comprises of random variables such as $\sigma^2_k$, $\mathbf{c}_\mathbf{w_{k}}$, $\mathbf{\theta}_k$, and $\mathbf{\eta_{k,t}}$, and the parameters of their corresponding distributions such as $\alpha$ and $\beta$, $\mathbf{\mu}_c$ and $\mathbf{\Sigma_c}$, $\mathbf{\mu}_\theta$ and $\mathbf{\Sigma_\theta}$, $\mathbf{\mu}_{\eta_k}$ and $\mathbf{\Sigma}_{\eta_k}$.
\end{definition}

\subsection{Re-sample Particles with Weights}
At time $t-1$, each arm $a^{(k)}$ maintains a fixed-size set of particles. We denote the particle set as $\mathcal{P}_{k,t-1}$ and assume the number of particles in $\mathcal{P}_{k,t-1}$ is $p$. Let $\mathcal{P}_{k,t-1}^{(i)}$ be the $i^{th}$ particles of  arm $a^{(k)}$ at time $t-1$, where $1\le i\le p$. Each particle $\mathcal{P}_{k,t-1}^{(i)}$ has a weight, denoted as $\rho^{(i)}$,  indicating its fitness for the new observed data at time $t$. Note that $\sum_{i=1}^{p}{\rho^{(i)}} = 1$. The fitness of each particle $\mathcal{P}_{k,t-1}^{(i)}$ is defined as the likelihood of the observed data $\mathbf{x}_t$ and $r_{k,t}$. Therefore,
\begin{equation}
  \rho^{(i)} \varpropto P(\mathbf{x}_t, r_{k,t}|\mathcal{P}_{k,t-1}^{(i)}).
\end{equation}
 Further, $y_{k,t}$ is the predicted value of $r_{k,t}$. The distribution of $y_{k,t}$, determined by $\mathbf{c}_{\mathbf{w}_k}$, $\theta_k$, $\sigma^2_k$ and $\eta_{k,t}$, is given in Equation~\ref{eg:genYNew}.
 Therefore, we can compute $\rho^{(i)}$ in proportional to the density value given $y_{k,t} = r_{k,t}$. Thus,
 \begin{equation*}
 \small
    \begin{split}
   \rho^{(i)} \varpropto \iint_{\eta_{k,t},\eta_{k,t-1}}\{\mathcal{N}(r_{k,t}|\mathbf{x}_t^\intercal(\mathbf{c}_{\mathbf{w}_k} + \theta_k\odot\eta_{k,t}),\sigma^2_k) \\ \mathcal{N}(\eta_{k,t}|\eta_{k,t-1},\mathcal{I}_d) \mathcal{N}(\eta_{k,t-1}|\mathbf{\mu}_{\eta_{k}},\mathbf{\Sigma}_{\eta_{k}})\} \\  d\eta_{k,t}\,d\eta_{k,t-1},
 \end{split}
 \end{equation*}
 where state variables $\eta_{k,t}$ and $\eta_{k,t-1}$ are integrated out due to their change over time, and $\mathbf{c}_{\mathbf{w}_k}$, $\theta_k$, $\sigma^2_k$ are from $\mathcal{P}_{k,t-1}^{(i)}$. Then we obtain
 \begin{equation}
 \small
    \label{eq:computeWeights}
    \rho^{(i)} \varpropto \mathcal{N}(\mathbf{m}_k,\mathbf{Q}_k),
 \end{equation}
 where
 \begin{equation}
 \small
 \begin{split}
   \mathbf{m}_k &= \mathbf{x}_t^\intercal(\mathbf{c}_{\mathbf{w}_k} + \theta_k\odot\mu_{\eta_{k}}),\\
   \mathbf{Q}_k &= \sigma^2_k+\mathbf{x}_t^\intercal\odot\theta_k(\mathcal{I}_d+\mathbf{\Sigma_{\eta_k}})\theta_k^\intercal\odot\mathbf{x}_t.
   \end{split}
   \label{eq:mAndQ}
 \end{equation}
 Before updating any parameters, a re-sampling process is conducted. We replace the particle set $\mathcal{P}_k$ with a new set $\mathcal{P'}_k$, where $\mathcal{P'}_k$ is generated from $\mathcal{P}_k$ using sampling with replacement based on the weights of particles. Then sequential parameter updating is based on $\mathcal{P'}_k.$

 \subsection{Latent State Inference}
 At time $t-1$, the sufficient statistics for state $\eta_{k,t-1}$ are the mean (i.e., $\mu_{\eta_k}$) and the covariance (i.e., $\mathbf{\Sigma}_{\eta_k}$). Provided with the new observation data $\mathbf{x}_t$ and $r_{k,t}$ at time $t$, the sufficient statistics for state $\eta_{k,t}$ need to be re-computed. We apply the Kalman filtering~\cite{harvey1990forecasting} method to recursively update the sufficient statistics for $\eta_{k,t}$ based on the new observation and the sufficient statistics at time $t-1$. Let $\mathbf{\mu'}_{\mathbf{\eta}_k}$ and $\mathbf{\Sigma'}_{\eta_k}$ be the new sufficient statistics of state $\eta_{k,t}$ at time $t$. Then,
 \begin{equation}
 \small
   \begin{split}
     &\mathbf{\mu'}_{\mathbf{\eta}_k} = \mathbf{\mu}_{\mathbf{\eta}_k} +\underbrace{\mathbf{G}_k(r_{k,t} - \mathbf{x}_t^\intercal(\mathbf{c}_{\mathbf{w}_k}+\theta_k\odot\eta_{k,t-1}))}_{\text{Correction by Kalman Gain}}, \\
     &\mathbf{\Sigma'}_{\eta_k} = \mathbf{\Sigma}_{\eta_k} + \mathcal{I}_d - \underbrace{\mathbf{G}_k\mathbf{Q}_k\mathbf{G}_k^\intercal}_{\text{Correction by Kalman Gain}},
   \end{split}
   \label{eq:stateUpdate}
 \end{equation}
 where $\mathbf{Q}_k$ is defined in Equation~\ref{eq:mAndQ} and $\mathbf{G}_k$ is Kalman Gain~\cite{harvey1990forecasting} defined as $$\mathbf{G}_k = (\mathcal{I}_d + \mathbf{\Sigma}_{\eta_k})\theta_k\odot\mathbf{x}_t\mathbf{Q}_k^{-1}.$$ As shown in Equation~\ref{eq:stateUpdate}, both $\mathbf{\mu'}_{\mathbf{\eta}_k}$ and $\mathbf{\Sigma'}_{\eta_k}$ are estimated with a correction using Kalman Gain $\mathbf{G}_k$(i.e., the last term in both two formulas).
 With the help of the sufficient statistics for the state random variable, $\eta_{k,t}$ can be draw from the Gaussian distribution
 \begin{equation}\label{eq:sampleState}
 \small
 \eta_{k,t} \thicksim \mathcal{N}(\mathbf{\mu'}_{\mathbf{\eta}_k},\mathbf{\Sigma'}_{\eta_k}).
 \end{equation}
 \subsection{Parameter Inference}
 At time $t-1$, the sufficient statistics for the parameter random variables ($\sigma^2_k$, $\mathbf{c}_\mathbf{w_{k}}$, $\mathbf{\theta}_k$) are ($\alpha$, $\beta$, $\mathbf{\mu}_c$, $\mathbf{\Sigma}_c$, $\mathbf{\mu}_\theta$, $\mathbf{\Sigma}_\theta$).

 Let $\mathbf{z}_t = (\mathbf{x}_t^\intercal, (\mathbf{x}_t\odot\mathbf{\eta_{k,t}})^\intercal)^\intercal$,
 $\mathbf{\Sigma} = \begin{bmatrix}
                        \mathbf{\Sigma_{c}} & \mathbf{0} \\
                        \mathbf{0} & \mathbf{\Sigma_{\theta}}
                    \end{bmatrix},$
 $\mathbf{\mu} = (\mathbf{\mu_{c}}^\intercal,\mathbf{\mu_{\theta}}^\intercal )^\intercal$, and
 $\mathbf{\nu_k} = (\mathbf{c_{w_k}}^\intercal,\mathbf{\theta_{k}}^\intercal )^\intercal$
 where $\mathbf{z}_t,$ $\mathbf{\mu}$, and $\mathbf{\nu}$ are $2d$-dimensional vector, $\mathbf{\Sigma}$ is a $2d\times 2d$ -dimensional matrix. Therefore, the inference of $\mathbf{c}_\mathbf{w_{k}}$ and $\mathbf{\theta}_k$ is equivalent to infer $\nu_k$ with its distribution $\nu_k \thicksim \mathcal{N}(\mu, \sigma^2_k\mathbf{\Sigma}).$ Assume $\mathbf{\Sigma'}$, $\mathbf{\mu'}$, $\alpha'$, and $\beta'$ be the sufficient statistics at time $t$ which are updated based on the sufficient statistics at time $t-1$ and the new observation data. The sufficient statistics for parameters are updated as follows:
 \begin{equation}
   \begin{split}
   \small
     &\mathbf{\Sigma'} = (\mathbf{\Sigma}^{-1} + \mathbf{z}_t\mathbf{z}_t^\intercal)^{-1},
     \ \ \ \mathbf{\mu'} = \mathbf{\Sigma'}(\mathbf{z}_tr_{k,t} + \mathbf{\Sigma}\mu), \\
     &\alpha' = \alpha + \frac{1}{2}, \\
     &\beta' = \beta+\frac{1}{2}(\mu^\intercal\mathbf{\Sigma}^{-1}\mu + r_{k,t}^2 - \mu'^\intercal\mathbf{\Sigma'}^{-1}\mu').
   \end{split}
   \label{eq:updateParameters}
 \end{equation}
 At time $t$, the sampling process for $\sigma^2_k$ and $\nu_k$ is summarized as follows:
  \begin{equation}
    \begin{split}
        &\sigma^2_k \thicksim \mathcal{IG}(\alpha',\beta'),
         \ \ \nu_k \thicksim \mathcal{N}(\mu',\sigma^2_k\mathbf{\Sigma'}).
    \end{split}
    \label{eq:sampleParameters}
  \end{equation}
  \subsection{Integration with Policies}\label{sec:policyIntegration}
  As discussed in Section \ref{sec:base}, both LinUCB and Thompson sampling allocate the pulling chance based on the posterior distribution of $\mathbf{w}_k$ and $\sigma^2_k$ with the hyper parameters $\mathbf{\mu_{w}}$, $\mathbf{\Sigma_{w}}$, $\alpha$, and $\beta.$

   As to the context drifting model, when $\mathbf{x}_t$ arrives at time $t$, the reward $r_{k,t}$ is unknown since it is not observed until one of arms is pulled.
   Without observed $r_{k,t}$, the particle re-sampling, latent state inference, and parameter inference for time $t$ can not be conducted.
   Furthermore, every arm has $p$ independent particles. Within each particle, the posterior distributions of $\mathbf{w}_{k,t-1}$ are not available since $\mathbf{w}_{k,t-1}$ has been decomposed into $\mathbf{c}_{\mathbf{w}_k}$, $\theta_k$, and $\eta_{k,t-1}$ based on Equation~\ref{eq:weights}.  We address these issues as follows.

   Within a single particle of arm $a^{(k)}$, the distribution of $\mathbf{w}_{k,t-1}$ can be derived by
   \begin{equation}
   \small
        \mathbf{w}_{k,t-1} \thicksim \mathcal{N}(\mathbf{\mu_{w_k}}, \sigma^2_k\mathbf{\Sigma_{w_k}}),
   \end{equation}
   where
   \begin{equation}
   \small
     \begin{split}
       &\mathbf{\mu_{w_k}} = \mu_{c} + (\mathbf{\Sigma}_{\eta_k} + \sigma^2_k\mathbf{\Sigma}_\theta)^{-1}(\mathbf{\Sigma}_{\eta_k}\mu_\theta + \sigma^2_k\mathbf{\Sigma_\theta}\mu_{\eta_k}), \\
       &\mathbf{\Sigma_{w_k}} = \sigma^2_k\mathbf{\Sigma}_c + \sigma^2_k\mathbf{\Sigma_\theta}\mathbf{\Sigma}_{\eta_k} (\mathbf{\Sigma}_{\eta_k} + \sigma^2_k\mathbf{\Sigma_\theta})^{-1}.
     \end{split}
   \end{equation}
   Let $\mathbf{w^{(i)}}_{k,t-1}$, $\mathbf{\mu^{(i)}_{w_k}}$, ${\sigma^2}^{(i)}_k$, and $\mathbf{\Sigma^{(i)}_{w_k}}$ be the random variables in the $i^{(th)}$ particle. We use the mean of $\mathbf{w}_{k,t-1}$, denoted as $\mathbf{\bar{w}}_{k,t-1}$, to infer the decision in the bandit algorithm. Therefore,
   \begin{equation}
   \small
    \label{eq:posterMean}
     \mathbf{\bar{w}}_{k,t-1} \thicksim \mathcal{N}(\mathbf{\bar{\mu}_{w_k}},\mathbf{\bar{\Sigma}_{w_k}}),
   \end{equation}
   where
   \begin{equation}
   \small
     \begin{split}
       &\mathbf{\bar{\mu}_{w_k}} = \frac{1}{p}\sum_{i=1}^{p}{\mathbf{\mu^{(i)}_{w_k}}},
        \ \ \ \mathbf{\bar{\Sigma}_{w_k}}= \frac{1}{p^2}\sum_{i=1}^{p}{{\sigma^2}^{(i)}_k\mathbf{\Sigma^{(i)}_{w_k}}}.
     \end{split}
   \end{equation}

By virtual of  Equation~\ref{eq:posterMean}, both Thompson sampling and LinUCB can address the bandit problem as mentioned in Section~\ref{sec:base}. Specifically, Thompson sampling draws $\mathbf{w}_{k,t}$ from Equation~\ref{eq:posterMean} and then predicts the reward for each arm with $\mathbf{w}_{k,t}$. The arm with maximum predicted reward is selected to pull. While LinUCB selects arm with a maximum score, where the score is defined as a combination of the expectation of $y_{k,t}$ and its standard deviation, i.e.,
 \begin{equation*}
 \small
   E(y_{k,t}|\mathbf{x}_t) + \lambda\sqrt{Var(y_{k,t}|\mathbf{x}_t)},
 \end{equation*} where $\lambda$ is predefined parameter, $E(y_{k,t}|\mathbf{x}_t)$ and $Var(y_{k,t}|\mathbf{x}_t)$ are computed by
 $$E(y_{k,t}|\mathbf{x}_t) = \mathbf{x}_t^\intercal\mathbf{w}_{k,t}. \ \ Var(y_{k,t}|\mathbf{x}_t)= \mathbf{x}_t^\intercal\mathbf{\bar{\Sigma}^{-1}_{w_k}}\mathbf{x}_t + \frac{1}{p^2}\sum_{i=1}^{p}{\sigma^2_k}.$$

  \subsection{Algorithm}
  Putting all the aforementioned things together, an algorithm based on the context drifting model is provided below.
  \begin{algorithm}[h!]
   \footnotesize
    \caption{The algorithm for context drift model (\texttt{Drift})}
     \label{alg:drift}
    \begin{algorithmic}[1]
        \Procedure{main}{$p$}\Comment{main entry}
            \State {Initialize arms with $p$ particles.}
            \For {$t\gets 1, T$}
            \State {Get $\mathbf{x}_t$.}
            \State {$a^{(k)} = \argmax_{j=1,K}{\text{EVAL}(a^{(j)},\mathbf{x}_t)}$}
            \State {Receive $r_{k,t}$ by pulling arm $a^{(k)}$.}
            \State {\text{UPDATE}($\mathbf{x}_t$, $a^{(k)}$, $r_{k,t}$). }
            \EndFor
        \EndProcedure
        \Statex{}
        \Procedure {eval}{$a^{(k)}$, $\mathbf{x}_t$}\Comment{get a score for $a^{(k)}$, given $\mathbf{x}_t.$}
            \State{Learn the parameters based on all particles' inferences of $a^{(k)}$ by Equation~\ref{eq:posterMean}.}
            \State{Compute a score based on the parameters learnt.}
            \State \textbf{return} {the score.}
        \EndProcedure
        \Statex{}
        \Procedure {update}{$\mathbf{x}_t$, $a^{(k)}$, $r_{k,t}$}\Comment{update the inference.}
            \For {$i\gets 1, p$}\Comment{Compute weights for each particle.}
                \State {Compute weight $\rho^{(i)}$ of particle $\mathcal{P}_k^{(i)}$ by Equation \ref{eq:computeWeights}.}
            \EndFor
            \State {Re-sample $\mathcal{P'}_k$ from $\mathcal{P}$ according to the weights $\rho^{(i)}$s.}
            \For {$i\gets 1, p$}\Comment{Update statistics for each particle.}
                \State {Update the sufficient statistics for $\eta_{k,t}$ by  Equation~\ref{eq:stateUpdate}.}
                \State {Sample $\eta_{k,t}$ according to Equation~\ref{eq:sampleState}.}
                \State {Update the statistics for $\sigma^2_k$, $\mathbf{c}_{\mathbf{w}_k}$, $\theta_k$ by Equation~\ref{eq:updateParameters}.}
                \State {Sample $\sigma^2_k$, $\mathbf{c}_{\mathbf{w}_k}$, $\theta_k$ according to Equation~\ref{eq:sampleParameters}.}
            \EndFor
        \EndProcedure
    \end{algorithmic}
  \end{algorithm}
Online inference for contextual multi-armed bandit problem starts with $\text{MAIN}$ procedure, as presented in Algorithm~\ref{alg:drift}. As $\mathbf{x}_t$ arrives at time $t$, the $\text{EVAL}$ procedure computes a score for each arm, where the definition of score depends on the specific policy. The arm with the highest score is selected to pull. After receiving a reward by pulling an arm, the new feedback is used to update the contextual drifting model by the $\text{UPDATE}$ procedure. Especially in the $\text{UPDATE}$ procedure, we use the $resample$-$propagate$ strategy in particle learning~\cite{carvalho2010particle} rather than the $propagate$-$resample$ strategy in particle filtering~\cite{djuric2003particle}. With the $resample$-$propagate$ strategy, the particles are re-sampled by taking $\rho^{(i)}$ as the $i^{th}$ particle's weight, where the $\rho^{(i)}$ indicates the occurring probability of the observation at time $t$ given the particle at time $t-1$. The $resample$-$propagate$ strategy is considered as an optimal and fully adapted strategy, avoiding an importance sampling step.
\subsection{Methodology to Hierarchial Dependency Modeling}
In this section, we propose the HMAB (\underline{H}ierarchical \underline{M}ulti-\underline{A}rmed \underline{B}andit) algorithms for exploiting the dependencies among arms organized hierarchically. 

At each time $t$, a policy $\pi$ will select a path $\pi_\mathcal{H}(\mathbf{x}_t|t)$ from $\mathcal{H}$ according to the context $\mathbf{x}_t$. Assuming $a^{(p)}\in\pi_\mathcal{H}(\mathbf{x}_t|t)$ is the leaf node (i.e., an item), then we have $pth(a^{(p)}) = \pi_\mathcal{H}(\mathbf{x}_t|t)$. After recommending item $a^{(p)}$, a reward $r_{p,t}$ is obtained. 
Since the reward $r_{p,t}$ is shared by all the arms along the path $pth(a^{(p)})$, a set of triples $F=\{(\mathbf{x}_t, a^{(k)}, r_{k,t})|a^{(k)}\in pth(a^{(k)}), r_{k, t}=r_{p, t}\}$ are acquired.
 A new sequence $S_{\pi,t}$ is generated by incorporating the triple set $F$ into $S_{\pi, t-1}$. The posterior distribution for every $a^{(k)}\in pth(a^{(k)})$ needs to be updated with the new feedback sequence $S_{\pi, t}$. The posterior distribution of $\mathbf{w}_{k}$ and $\sigma^2_{k}$ given $S_{\pi,t}$ is a $\mathcal{NIG}$ distribution with the hyper parameter $\mathbf{\mu_w}$, $\mathbf{\Sigma_w}$, $\alpha_{k}$ and $\beta_{k}$. These hyper parameters at time $t$ are updated based on their values at time $t-1$ according to Equation~\ref{eq:posteiorUpdate}.

 \begin{algorithm}[h!]
    \footnotesize
    \caption{The algorithm for hierarchical dependency model}
     \label{alg:hmab}
    \begin{algorithmic}[1]
            \Procedure{main}{$\mathcal{H}, \pi, \lambda$}\Comment{main entry, $\pi$ is the policy}
            \For {$t \leftarrow 1,T$}
                \State {Initialize parameters of  $a^{(m)}~\in~\mathcal{H}$ to $\alpha$, $\beta$, $\mathbf{\Sigma}_{\mathbf{w}}~=~\mathbf{I}_d$, $\mathbf{\mu}_{\mathbf{w}}~=~\mathbf{0}_{d \times 1}$.}
                \State {Get contextual vector $\mathbf{x}_{t} \in \mathcal{X}$.}
                \For {each path $P$ of $\mathcal{H}$}
                    \State Compute the reward of $P$ using Equation~\ref{eq:one-time-reward}, by calling $\text{EVAL}(\mathbf{x}_t, a^{(k)}, \pi)$ for each arm $a^{(k)}\in P$.
                \EndFor
                \State {Choose the path $P^*$ with maximum reward.}
                \State {Recommend item $a^{(*)}$ (leaf node of $P^*$).}
                \State {Receive reward $r_{*,t}$ by pulling arm $a^{(*)}$.}
                \State {$\text{UPDATE}(\mathbf{x}_t, P^{*}, r_{*,t}, \pi)$.}
            \EndFor
            \EndProcedure
            \State {}
        \Procedure {eval}{$\mathbf{x}_t$, $a^{(k)}$, $\pi$}\Comment{get a score for $a^{(k)}$, given $\mathbf{x}_t$}
               \If {$\pi$ is TS}
                \State {Sample $\sigma^2_{k,t}$, $\mathbf{w}_{k,t}$ according to Equation~\ref{eq:nig}.}
                \State {\textbf{return} $y_{k,t} = \mathbf{x}^{T}_{t}$$\mathbf{w}_{k,t}$.}
               \EndIf
               \If {$\pi$ is LinUCB}
                \State {\textbf{return} $y_{k,t} = \mathbf{x}^{T}_{t}$$\mathbf{\mu}_{\mathbf{w}_{t-1}} + \frac{\lambda}{\sigma_{t-1}} \sqrt{\mathbf{x}^{T}_t \mathbf{\Sigma}^{-1}_{\mathbf{w}_{t-1}}\mathbf{x}_t}$.}
               \EndIf
        \EndProcedure
        \State {}
        \Procedure {update}{$\mathbf{x}_t, P, r_t, \pi$}\Comment{update the inference.}
            \Comment {$P$ is the path in $\mathcal{H}$, $r_{t}$ is the reward.}
            \For {each arm $a^{(k)} \in P$}
            \State {Update $\alpha$, $\beta$, $\mathbf{\Sigma}_{\mathbf{w}_{t}}$, $\mathbf{\mu}_{\mathbf{w}_{t}}$ using Equation~\ref{eq:posteiorUpdate}.}
            \EndFor

        \EndProcedure
    \end{algorithmic}
  \end{algorithm}

Note that the posterior distribution of $\mathbf{w}_k$ and $\sigma^2_k$ at time $t$ is considered as the prior distribution of time $t+1$. On the basis of the aforementioned inference of the leaf node $a^{(k)}$, we propose HMAB algorithms presented in Algorithm~\ref{alg:hmab} developing different strategies including HMAB-TS($\mathcal{H}$, $\alpha$, $\beta$) and HMAB-LinUCB($\mathcal{H}$, $\lambda$).

\section{Experiment Setup}\label{sec:experiment}
To demonstrate the efficacy of our proposed model, extensive experiments are conducted on three real-world datasets. The KDD Cup 2012~\footnote{http://www.kddcup2012.org/c/kddcup2012-track2} for online advertising recommendation and Yahoo! Today News for online news recommendation are used to verify our context drift model, while IT ticket dataset for automation recommendation collected by IBM Tivoli Monitoring system~\footnote{http://ibm.com/software/tivoli/} for the hierarchical dependency model.
We first describe the dataset and evaluation method.
Then we discuss the comparative experimental results of the proposed and baseline algorithms.

\subsection{Baseline Algorithms}\label{sec:baseline}
In the experiment, we demonstrate the performance of our method by comparing with the following algorithms. The baseline algorithms include:
\vspace{-0.1in}
\begin{enumerate}
 \itemsep-0.05in
  \item \texttt{Random}: it randomly selects an arm to pull without considering any contextual information.
  \item \texttt{$\epsilon$-greedy}($\epsilon$) (or \textit{EPSgreedy}): it randomly selects an arm with probability $\epsilon$ and selects the arm of the largest predicted reward with probability $1-\epsilon$, where $\epsilon$ is a predefined parameter. When $\epsilon = 0$, it is equivalent to the \texttt{Exploit} policy.
  \item \texttt{GenUCB}($\lambda$): it denotes the general \texttt{UCB} algorithm for contextual bandit problems. It can be integrated with linear regression model (e.g.,\texttt{LinUCB}~\cite{li2010contextual}) or logistic regression model for reward prediction. Both \texttt{LinUCB} and \texttt{LogUCB} take the parameter $\lambda$ to obtain a score defined as a linear combination of the expectation and the deviation of the reward.
  \item \texttt{TS}($q_0$): Thompson sampling described in Section~\ref{sec:base}, randomly draws the coefficients from the posterior distribution, and selects the arm of the largest predicted reward. The priori distribution is $\mathcal{N}(\mathbf{0}, q_0^{-1}\mathbf{I})$.
  \item \texttt{TSNR}($q_0$): it is similar to \texttt{TS}($q_0$), but in the stochastic gradient ascent, there is no regularization by the prior. The priori distribution $\mathcal{N}(\mathbf{0}, q_0^{-1}\mathbf{I})$ is only used in the calculation of the posterior distribution for the parameter sampling, but not in the stochastic gradient ascent. When $q_0$ is arbitrarily large, the variance approaches 0 and \texttt{TSNR} becomes \texttt{Exploit}.
  \item \texttt{Bootstrap}: it is non-Bayesian but an ensemble method for arm selection. Basically, it maintains a set of bootstrap samples for each arm and randomly pick one bootstrap sample for inference~\cite{tang2015personalized}.
\end{enumerate}

Our methods proposed in this paper include:
\begin{enumerate}
 \itemsep-0.05in
  \item \texttt{TVUCB}($\lambda$): it denotes the time varying \texttt{UCB} which integrates our proposed context drift model with \texttt{UCB} bandit algorithm. Similar to \texttt{LinUCB}, the parameter $\lambda$ is given.
  \item \texttt{TVTP}($q_0$): it denotes the time varying \texttt{Thompson sampling} algorithm which is extended with our proposed context drift model and the algorithm is outlined in Algorithm~\ref{alg:drift}. The parameter $q_0$, similar to \texttt{TS}($q_0$), specifies the prior distribution of the coefficients.
  \item \texttt{HAMB-EpsGreedy}($\mathcal{H}$, $\epsilon$): a random arm with probability $\epsilon$ is selected, and the arm of the highest estimated reward $\hat{r}_{k,t}$ with probability $1-\epsilon$ with respect to the hierarchy $\mathcal{H}$, which is a predefined parameter as well as $\epsilon$.
 	\item \texttt{HMAB-TS}($\mathcal{H}, \alpha, \beta$): it denotes our proposed hierarchical multi-armed bandit with \texttt{Thompson sampling} outlined in Algorithm~\ref{alg:hmab}. $\mathcal{H}$ is the taxonomy defined by domain experts. $\alpha$ and $\beta$ are hyper parameters.
 	\item \texttt{HMAB-LinUCB}($\mathcal{H}, \lambda$): it represents our proposed algorithm based on \texttt{LinUCB} presented in Algorithm~\ref{alg:hmab}. Similarly, $\mathcal{H}$ is the hierarchy depicting the dependencies among arms. And the parameter $\lambda$ is given with the same use in \texttt{LinUCB}.
\end{enumerate}

\subsection{KDD Cup 2012 Online Advertising}
\subsubsection{Description}
  Online advertising has become one of the major revenue sources of the Internet industry for many years. In order to maximize the Click-Though Rate (CTR) of displayed advertisements (ads), online advertising systems need to deliver these appropriate ads to individual users. Given the context information, sponsored search which is one type of online advertising will display a recommended ad in the search result page. Practically, an enormous amount of new ads will be continuously brought into the ad pool. These new ads have to be displayed to users, and feedbacks have to be collected for improving the system's CTR prediction. Thereby, the problem of ad recommendation can be regarded as an instance of contextual bandit problem. In this problem, an arm is an ad, a pull is an ad impression for a search activity, the context is the information vector of user profile and search keywords, and the reward is the feedbacks of user's click on ads.
  \subsubsection{Evaluation Method}
  We first use a simulation method to evaluate the KDD Cup 2012 online ads data, which is applied in~\cite{chapelle2011empirical} as well. The simulation and \emph{replayer}~\cite{li2012unbiased} are two of the frequently used methods for the bandit problem evaluation. As discussed in \cite{chapelle2011empirical} and \cite{tang2015personalized}, the simulation method performs better than \emph{replayer} method when the item pool contains a large number of recommending items, especially larger than 50. The large number of recommending items leads to the CTR estimation with a large variance due to the small number of matched visits.

  In this data set, we build our ads pool by randomly selecting $K = 100$ ads from the entire set of ads. There is no explicit time stamp associated with each ad impression, and we assume the ad impression arrives in chronological order with a single time unit interval between two adjacent impressions. The context information of these ads are real and obtained from the given data set. However, the reward of the $k^{th}$ ad is simulated with a coefficient vector $\mathbf{w}_{k,t}$, which dynamically changes over time. Let $\varrho$ be the change probability, where each coefficient keeps unchanged with probability $1-\varrho$ and varies dynamically with probability $\varrho$. We model the dynamical change as a Gaussian random walk by $\mathbf{w}_{k,t} = \mathbf{w}_{k,t} +\Delta_w$ where $\Delta_w$ follows the standard Gaussian distribution, i.e., $\Delta_w \thicksim \mathcal{N}(\mathbf{0}, \mathcal{I}_d)$. Given a context vector $\mathbf{x}_t$ at time $t$, the click of the $k^{th}$ ad is generated with a probability $(1+\exp(-\mathbf{w}_{k,t}^T\mathbf{x}_t))^{-1}.$ For each user visit and each arm, the initial weight vector $\mathbf{w}_{k,0}$ is drawn from a fixed normal distribution that is randomly generated before the evaluation.

\subsubsection{Context Change Tracking}
With the help of the simulation method, we get a chance to know the ground truth of the coefficients. Therefore, we first explore the fitness of our model with respect to the true coefficient values over time. Then we conduct our experiment over the whole online ads data set containing 1 million impressions by using the CTR as the evaluation metric.

\begin{figure}[h]
\centering
   \scalebox{0.35}{
    \epsfig{file=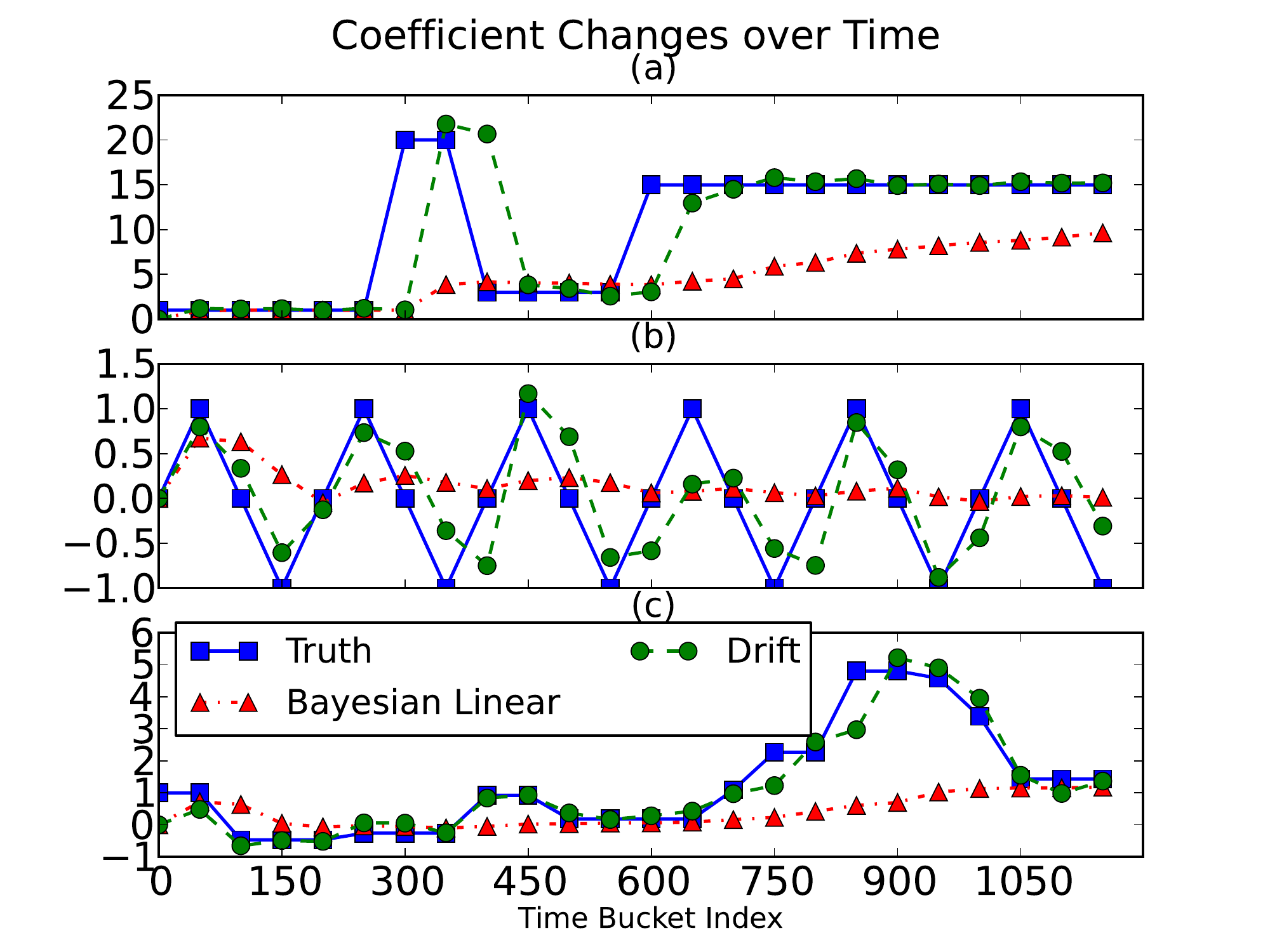}
    }
    \vspace{-0.1in}
    \caption{A segment of data originated from the whole data set is provided. The reward is simulated by choosing one dimension of the coefficient vector, which is assumed to vary over time in three different ways. Each time bucket contains $100$ time units.}
    \label{fig:tv-coef-simulation}
\vspace{-0.1in}
\end{figure}

 We simulate the dynamical change of coefficients in multiple different ways including the random walk over a small segment of data set shown in Figure~\ref{fig:tv-coef-simulation} from our previous work~\cite{zeng2016online}. Sampling a segment of data containing $120k$ impressions from the whole data set, we assume a dynamical change occurring on only one dimension of the coefficient vector, keeping other dimensions constant.  In $(a)$, we divide the whole segment of data into four intervals, where each has a different coefficient value. In $(b)$, we assume the coefficient value of the dimension changes periodically. In $(c)$, a random walk mentioned above is assumed, where $\varrho=0.0001.$ We compare our algorithm \texttt{Drift} with the bandit algorithm such as \texttt{LinUCB} with Bayesian linear regression for reward prediction. We set \texttt{Drift} with $5$ particles. It shows that our algorithm can fit the coefficients better than Bayesian linear regression and  can adaptively capture the dynamical change instantly. The reason is that, \texttt{Drift} has a random walk for each particle at each time and estimates the coefficient by re-sampling these particles according to their goodness of fitting.

\subsubsection{CTR Optimization for Online Ads }\label{sec:kddcup-ctr}
\begin{figure}[h!]
\vspace{-0.05in}
    \centering
    \scalebox{0.35}{
    \epsfig{file=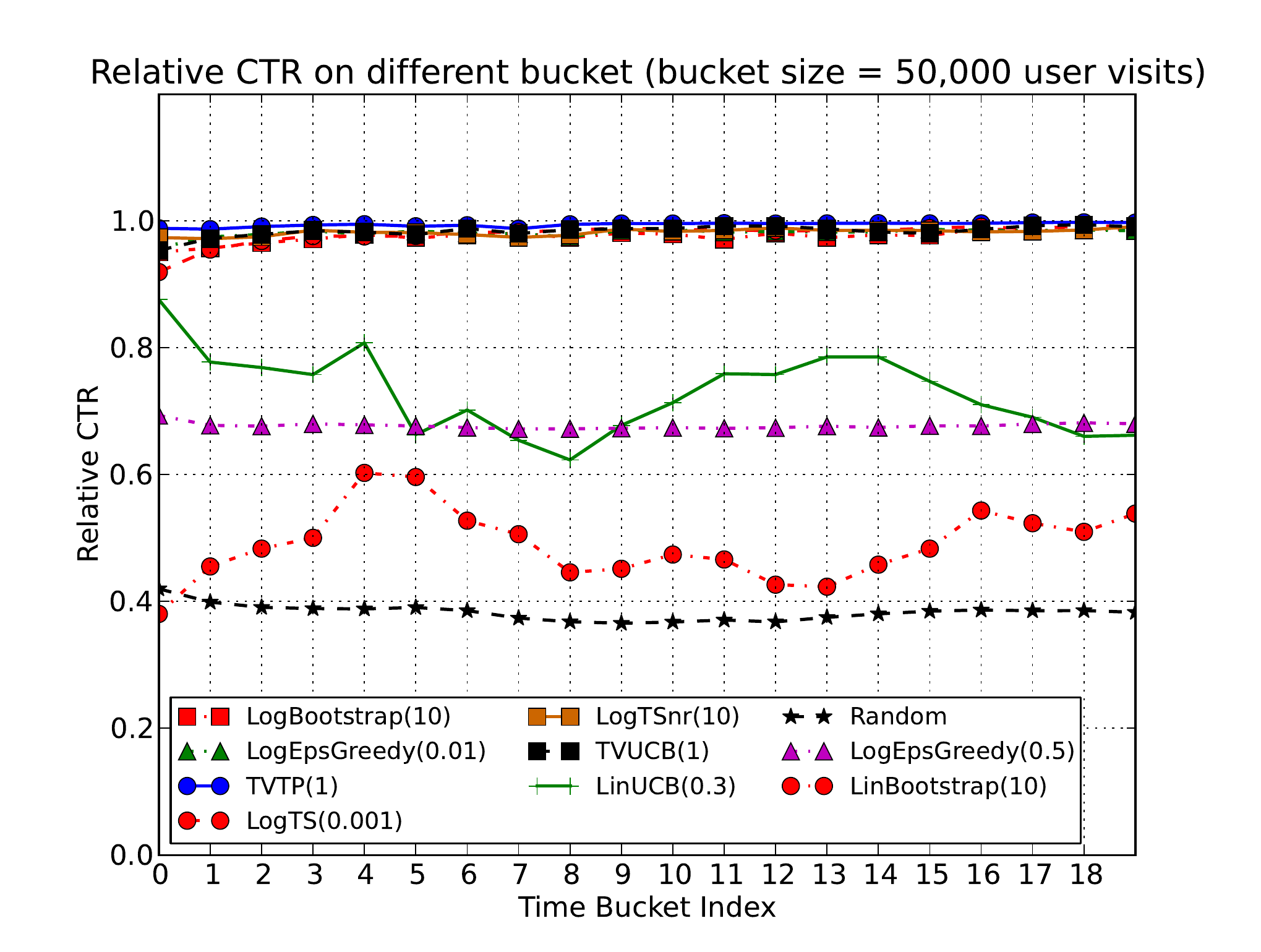}
    }
    \vspace{-0.1in}
    \caption{The CTR of KDD CUP 2012 online ads data is given for each time bucket. \texttt{LogBooststrap}, \texttt{LogTS}, \texttt{LogTSnr}, and \texttt{LogEpsGreedy} are bandit algorithms with logistic regression model. \texttt{LinUCB}, \texttt{LinBoostrap}, \texttt{TVTP}, and \texttt{TVUCB} are based on linear regression model.}
    \label{fig:ctr-kdd-bucket}
\vspace{-0.1in}
\end{figure}

In this section, we evaluate our algorithm over the online ads data in terms of CTR. The performance of each baseline algorithm listed in Section~\ref{sec:baseline} depends on the underlying reward prediction model (e.g., logistic regression, linear regression) and its corresponding parameters. 
Therefore, we first conduct the performance comparison for each algorithm with different reward prediction models and diverse parameter settings. Then the one with best performance is selected to compare with our proposed algorithm. 
The experimental results are presented in Figure~\ref{fig:ctr-kdd-bucket}~\cite{zeng2016online}. The algorithm $\texttt{LogBoostrap(10)}$ achieves better performance than \texttt{LinBootstrap(10)} since our simulation method is based on the $Logit$ function.

Although our algorithms \texttt{TVTP(1)} and \texttt{TVUCB(1)} are based on linear regression model, they can still achieve high CTRs and their performance is comparable to those algorithms based on logistic regression method such as, \texttt{LogTS(0.001)},\texttt{LogTSnr(10)}. The reason is that both \texttt{TVTP} and \texttt{TVUCB} are capable of capturing the non-linear reward mapping function by explicitly considering the context drift. The algorithm \texttt{LogEpsGreedy(0.5)} does not perform well. The reason is that the value of parameter $\epsilon$ is large, incurring lots of exploration.

\subsection{Yahoo! Today News}
 \subsubsection{Description}
  The core task of personalized news recommendation is to display appropriate news articles on the web page for the users according to the potential interests of individuals.  However, it is difficult to track the dynamical interests of users only based on the content. Therefore, the recommender system often takes the instant feedbacks from users into account to improve the prediction of the potential interests of individuals, where the user feedbacks are about whether the users click the recommended article or not. Additionally, every news article does not receive any feedbacks unless the news article is displayed to the user. Accordingly, we formulate the personalized news recommendation problem as an instance of contextual multi-arm bandit problem, where each arm corresponds to a news article and the contextual information including both content and user information.
 \subsubsection{Evaluation Method}
  We apply the \emph{replayer} method to evaluate our proposal method on the news data collection since the number of articles in the pool is not larger than $50$. The \emph{replayer} method is first introduced in~\cite{li2012unbiased}, which provides an unbiased offline evaluation via the historical logs. The main idea of \emph{replayer} is to replay each user visit to the algorithm under evaluation. If the recommended article by the testing algorithm is identical to the one in the historical log, this visit is considered as an impression of this article to the user. The ratio between the number of user clicks and the number of impressions is referred as CTR. The work in~\cite{li2012unbiased} shows that the CTR estimated by the \emph{replayer} method approaches the real CTR of the deployed online system if the items in historical user visits are randomly recommended.

  \subsubsection{CTR Optimization for News Recommendation}
  \begin{figure}[h]
    \vspace{-0.1in}
    \centering
    \scalebox{0.35}{
    \epsfig{file=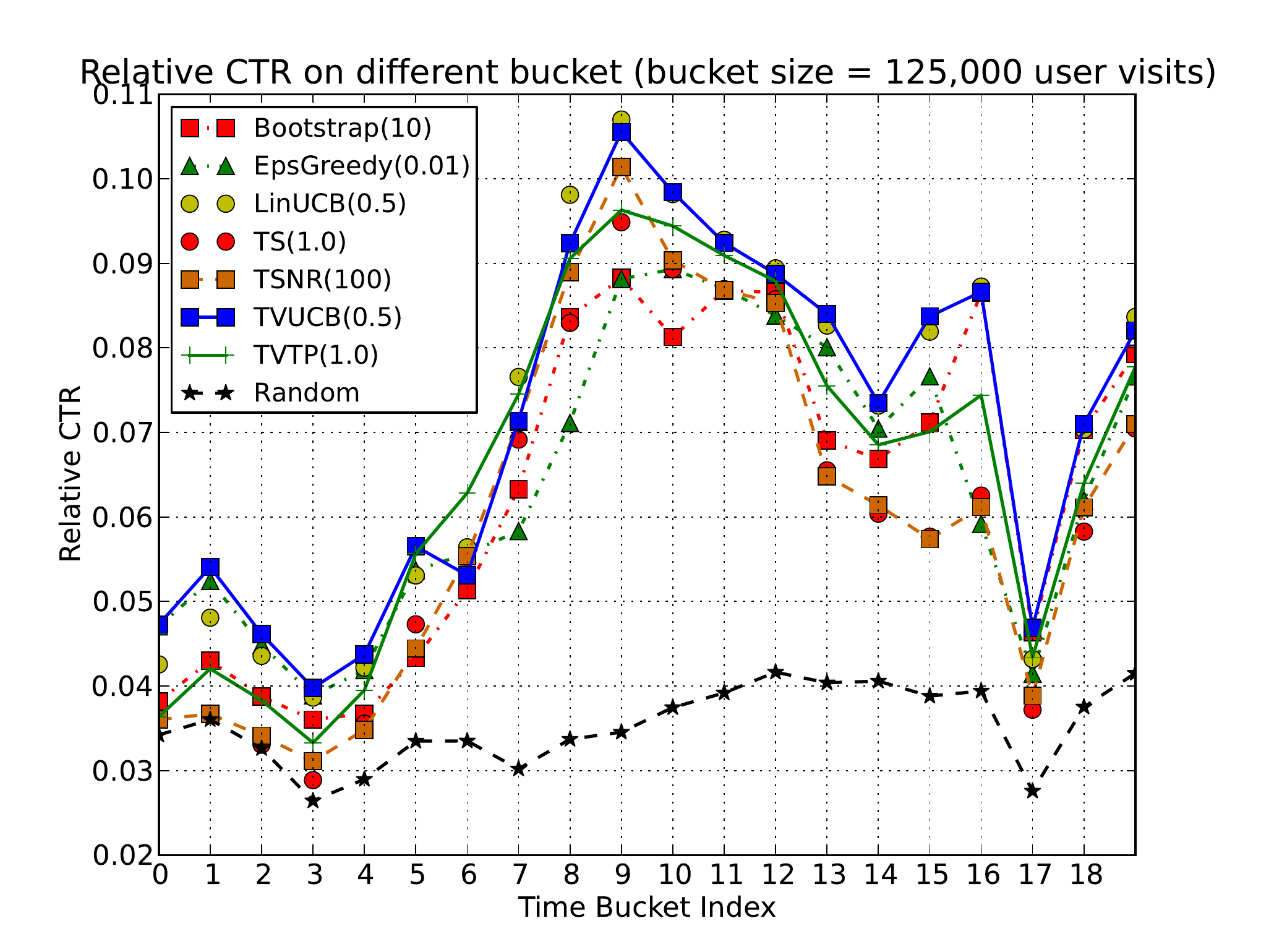}
    }
    \vspace{-0.15in}
    \caption{The CTR of Yahoo! News data is given for each time bucket. Those baseline algorithms are configured with their best parameters settings.}
    \label{fig:ctr-ydata-distribution}
\vspace{-0.1in}
\end{figure}
  Similar to the CTR optimization for online ads data, we conduct the performance comparison on different time buckets in Figure~\ref{fig:ctr-ydata-distribution} from ~\cite{zeng2016online}, where each algorithm is configured with the setting with the highest reward. The algorithm \texttt{TVUCB(0.5)} and \texttt{EpsGreedy(0.01)} outperforms others among the first four buckets, known as cold-start phrase when the algorithms are not trained with sufficient observations. After the fourth bucket, the performance of both \texttt{TVUCB(0.5)} and \texttt{LinUCB(0.5)} constantly exceeds the ones of other algorithms. In general, \texttt{TVTP(1.0)} performs better than \texttt{TS(1.0)} and \texttt{TSNR(100)}, where all the three algorithms are based on the \texttt{Thompson} \texttt{sampling}. Overall, \\ \texttt{TVUCB(0.5)} consistently achieves the best performance.

\subsection{IBM Global IT Ticket Dataset}
\subsubsection{Description}
 The increasing complexity of IT environments urgently requires the use of analytical approaches and automated problem resolution for more efficient delivery of IT services~\cite{zeng2014hierarchical,zeng2017knowledge,tang2017integrated,wang2017constructing,zhou2017star}. The core task of IT automation services is to automatically execute a recommended automation (i.e., a scripted resolution) to fix the current alert key (i.e., ticket problem) and the interactive feedback (e.g., success or failure) is used for continuous enhancements. Domain experts would define the hierarchical taxonomy to categorize IT problems, while these automations have corresponding categories. To utilize the taxonomy, we formulate it as a contextual bandit problem with dependent arms in the form of hierarchies.

\begin{figure}[htp!]
    \vspace{-0.05in}
    \centering
    \captionsetup{font=small}
    \scalebox{0.42}{
    \includegraphics{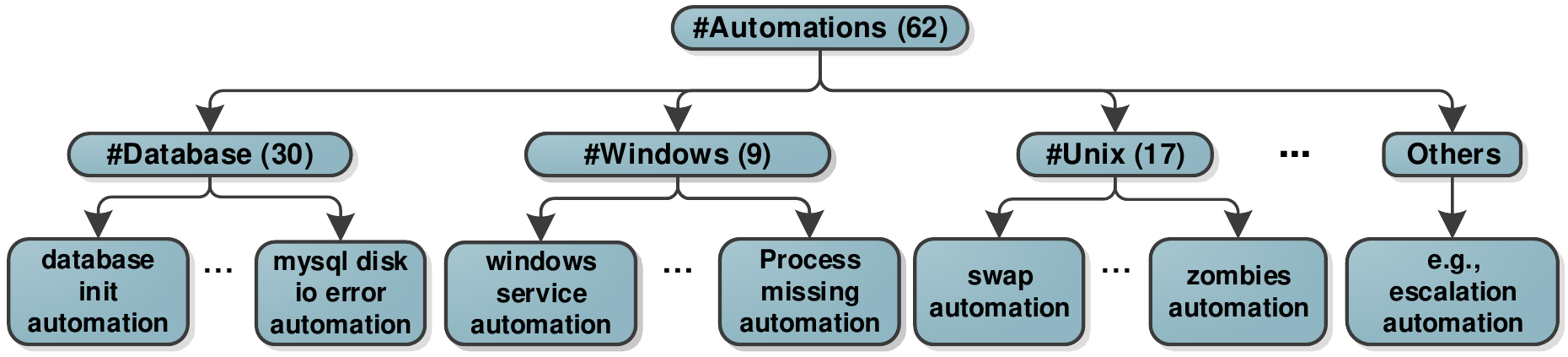}
    }
    \vspace{-0.05in}
    \caption{An automation hierarchy defined by domain experts.}
    \label{fig:automation-hierarchy}
   \vspace{-0.1in}
\end{figure}

 The dataset is collected by IBM Tivoli Monitoring system from July 2016 to March 2017, which contains 332,211 historical resolution records, which contains 1,091 alert keys (e.g., cpusum\_xuxc\_aix, prccpu\_rlzc\_std) and 62 automations (e.g., NFS automation, process CPU spike automation) in total. The execution feedback (i.e., reward or rating) indicating whether the ticket has been resolved by an automation or needs to be escalated to human engineers, is collected and utilized for our proposed model inference. Each record is stamped with the reporting time of the ticket. Additionally, a three-layer hierarchy $\mathcal{H}$ (see Figure~\ref{fig:automation-hierarchy}) given by domain experts is introduced to depict the dependencies among automations. The evaluation Method is the same used in Yahoo! News Today.

\subsubsection{Relative Success Rate Optimization}
We use the success rate as the evaluation metric in our experiments.
The higher success rate, the better performance of the algorithm. To avoid the leakage of business-sensitive information,  RSR (\underline{r}elative \underline{s}uccess \underline{r}ate) is reported. The RSR is the overall success rate of an algorithm divided by the overall success rate of random selection method where an automation is randomly recommended for ticket resolving. The following outlines the experimental performance of HMAB.

\begin{figure*}[htp!]
\captionsetup{font=small}
\scalebox{0.7}{
\begin{subfigure}{0.45\textwidth}
\centering
  \captionsetup{font=small}
  \scalebox{0.22}{
   \includegraphics{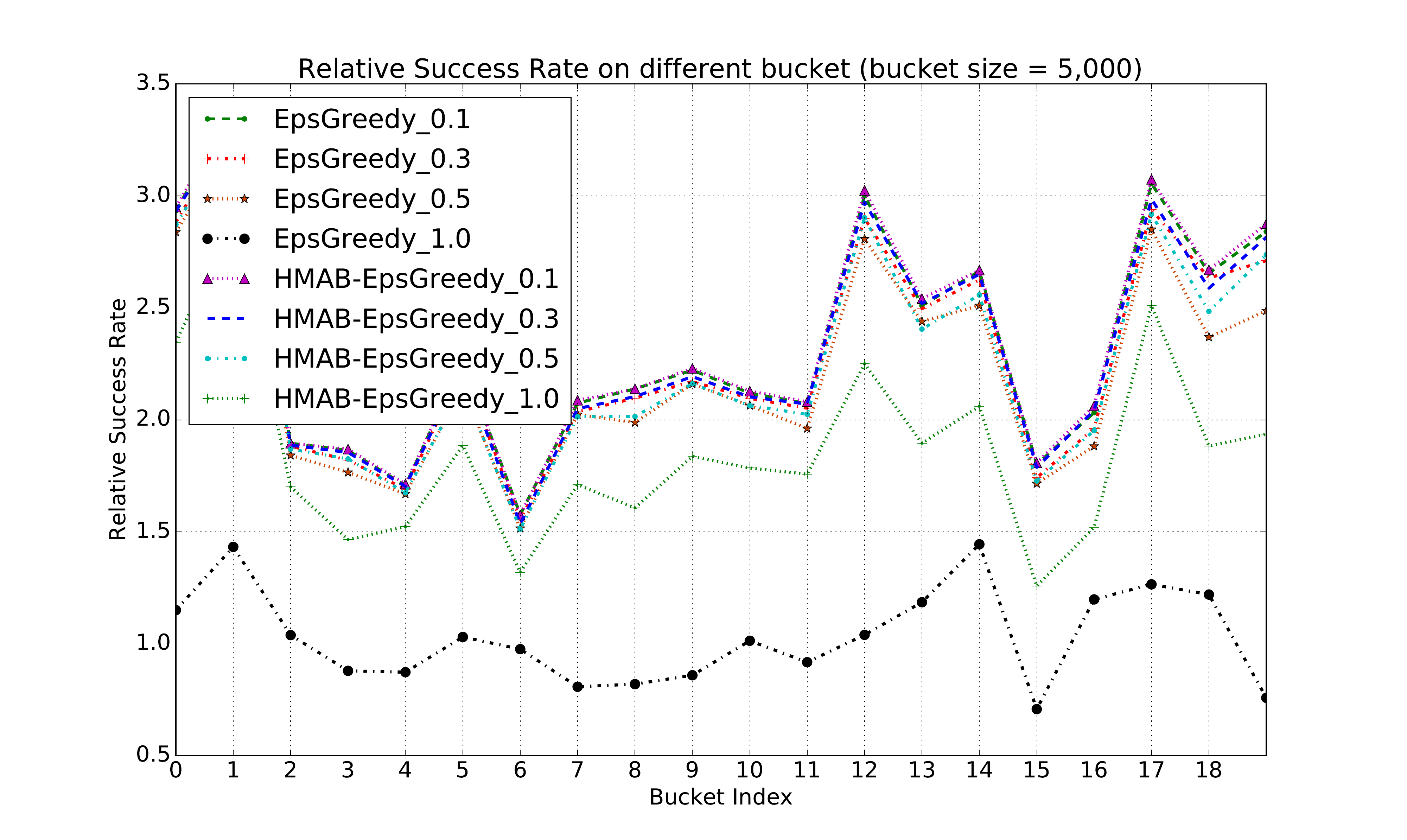}
  }
  \vspace{-0.15in}
\caption{$\epsilon$-greedy and HMAB-$\epsilon$-greedy.}
\label{fig:epsilon-performance}
\end{subfigure}

\begin{subfigure}{0.45\textwidth}
 \captionsetup{font=small}
  \scalebox{0.22}{
   \includegraphics{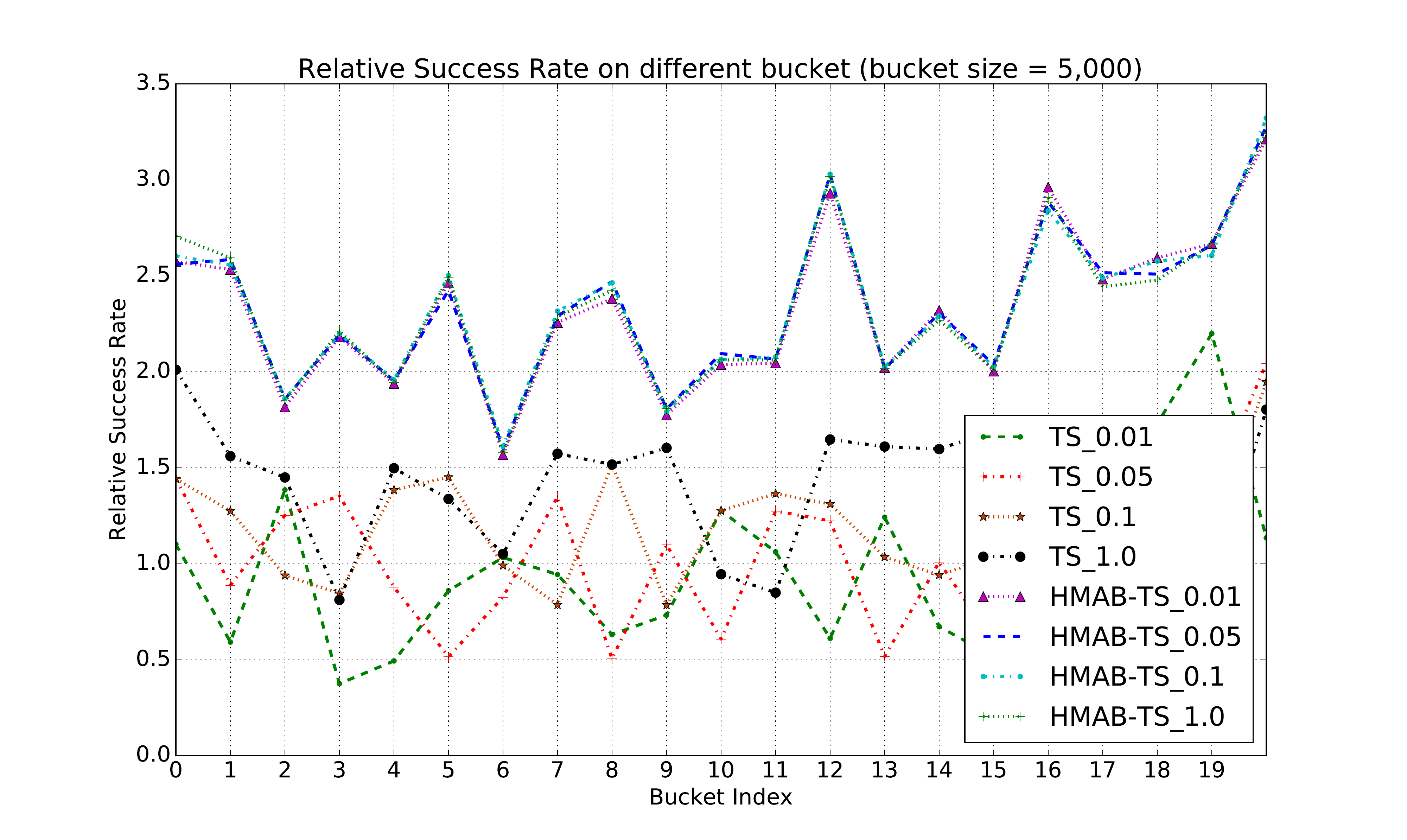}
  }
  \vspace{-0.15in}
\caption{TS and HMAB-TS.}
\label{fig:ts-performance}
\end{subfigure}

\begin{subfigure}{0.45\textwidth}
 \captionsetup{font=small}
  \scalebox{0.22}{
   \includegraphics{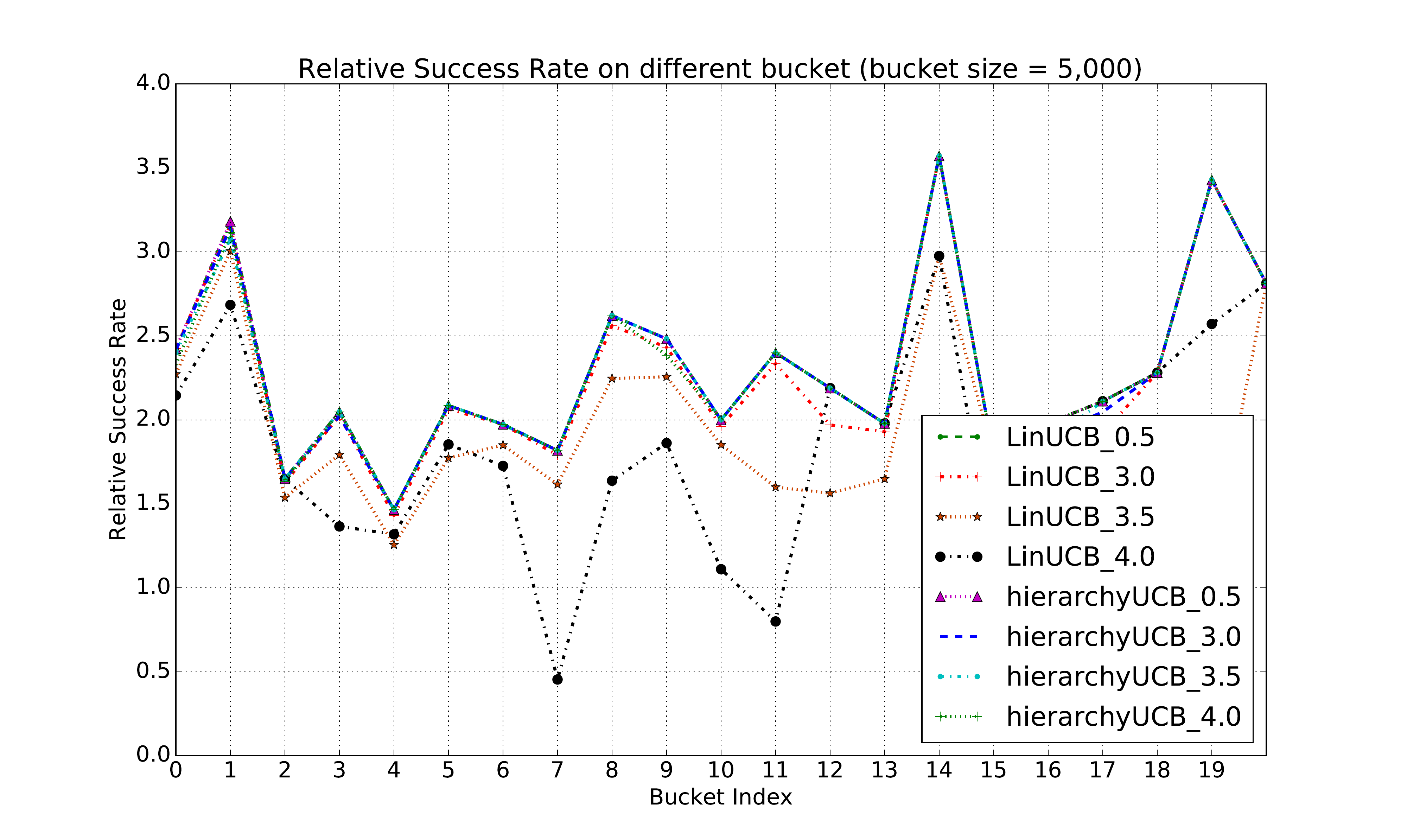}
  }
  \vspace{-0.15in}
\caption{LinUCB and HMAB-LinUCB.}
\label{fig:ucb-performance}
\end{subfigure}
}
\vspace{-0.1in}
\caption{The RSR of proposed and baseline algorithms on the dataset is given along different time buckets with diverse parameter settings.}
\label{fig:rsr-performance}
\vspace{-0.15in}
\end{figure*}

We demonstrate the performance of our proposed algorithms by comparing with the baseline algorithms including $\epsilon$-greedy~\cite{tokic2010adaptive}, Thompson sampling~\cite{chapelle2011empirical}, UCB~\cite{auer2002using}, and LinUCB~\cite{li2010contextual}).
Figure~\ref{fig:epsilon-performance}, Figure~\ref{fig:ts-performance}, and Figure~\ref{fig:ucb-performance}
 show the performance comparison between HMABs and the corresponding baselines configured with different parameter settings.
To clarify, we only list the performance for LinUCB and HMAB-LinUCB with the parameter $\lambda > 1$ in~Figure~\ref{fig:ucb-performance} to reveal the merits of HMAB-LinUCB since both algorithms perform similarly when $\lambda < 1$. By observing the experimental results, HMABs performs much better than the baselines and HMAB-LinUCB outperforms all other algorithms.

\section{Conclusion} \label{sec:conclusion}
The explosive growth of online services has driven many Internet companies to develop intelligent interactive recommendation services. In this paper, we first identify the preliminary challenges for modern recommender systems. Two novel and generic models are proposed to deal with these challenges, which have been verified on three practical real-world applications.

Deep reinforcement learning~\cite{arulkumaran2017brief}, trying to build automatic systems with a higher level understanding of the dynamic world, is poised to revolutionise the field of AI. In~\cite{zheng2018drn}, a deep reinforcement learning framework is proposed for news recommendation. Therefore, it is interesting to design the deep bandit framework~\cite{riquelme2018deep} for the future recommender systems.


\bibliographystyle{abbrv} 
\bibliography{bibs4cacm}  

\begin{thebibliography}{10}

\bibitem{adomavicius2005toward}
G.~Adomavicius and A.~Tuzhilin.
\newblock Toward the next generation of recommender systems: A survey of the
  state-of-the-art and possible extensions.
\newblock {\em IEEE transactions on knowledge and data engineering},
  17(6):734--749, 2005.

\bibitem{agrawal2013thompson}
S.~Agrawal and N.~Goyal.
\newblock Thompson sampling for contextual bandits with linear payoffs.
\newblock In {\em ICML}, pages 127--135, 2013.

\bibitem{arulkumaran2017brief}
K.~Arulkumaran, M.~P. Deisenroth, M.~Brundage, and A.~A. Bharath.
\newblock A brief survey of deep reinforcement learning.
\newblock {\em arXiv preprint arXiv:1708.05866}, 2017.

\bibitem{auer2002using}
P.~Auer.
\newblock Using confidence bounds for exploitation-exploration trade-offs.
\newblock {\em JMLR}, 3(Nov):397--422, 2002.

\bibitem{carvalho2010particle}
C.~Carvalho, M.~S. Johannes, H.~F. Lopes, and N.~Polson.
\newblock Particle learning and smoothing.
\newblock {\em Statistical Science}, 25(1):88--106, 2010.

\bibitem{chang2015space}
S.~Chang, J.~Zhou, P.~Chubak, J.~Hu, and T.~S. Huang.
\newblock A space alignment method for cold-start tv show recommendations.
\newblock In {\em Proceedings of the 24th International Conference on
  Artificial Intelligence}, pages 3373--3379. AAAI Press, 2015.

\bibitem{chapelle2011empirical}
O.~Chapelle and L.~Li.
\newblock An empirical evaluation of thompson sampling.
\newblock In {\em NIPS}, pages 2249--2257, 2011.

\bibitem{djuric2003particle}
P.~M. Djuri{\'c}, J.~H. Kotecha, J.~Zhang, Y.~Huang, T.~Ghirmai, M.~F. Bugallo,
  and J.~Miguez.
\newblock Particle filtering.
\newblock {\em Signal Processing Magazine, IEEE}, 20(5):19--38, 2003.

\bibitem{doucet2000sequential}
A.~Doucet, S.~Godsill, and C.~Andrieu.
\newblock On sequential monte carlo sampling methods for bayesian filtering.
\newblock {\em Statistics and computing}, 10(3):197--208, 2000.

\bibitem{gentile2014online}
C.~Gentile, S.~Li, and G.~Zappella.
\newblock Online clustering of bandits.
\newblock In {\em ICML}, pages 757--765, 2014.

\bibitem{halton1962sequential}
J.~H. Halton.
\newblock Sequential monte carlo.
\newblock In {\em Mathematical Proceedings of the Cambridge Philosophical
  Society}, volume~58, pages 57--78. Cambridge Univ Press, 1962.

\bibitem{harvey1990forecasting}
A.~C. Harvey.
\newblock {\em Forecasting, structural time series models and the Kalman
  filter}.
\newblock Cambridge university press, 1990.

\bibitem{hill2017efficient}
D.~N. Hill, H.~Nassif, Y.~Liu, A.~Iyer, and S.~Vishwanathan.
\newblock An efficient bandit algorithm for realtime multivariate optimization.
\newblock In {\em SIGKDD}, pages 1813--1821. ACM, 2017.

\bibitem{hu2008collaborative}
Y.~Hu, Y.~Koren, and C.~Volinsky.
\newblock Collaborative filtering for implicit feedback datasets.
\newblock In {\em ICDM}, pages 263--272. Ieee, 2008.

\bibitem{jannach2016recommender}
D.~Jannach, P.~Resnick, A.~Tuzhilin, and M.~Zanker.
\newblock Recommender systems beyond matrix completion.
\newblock {\em Communications of the ACM}, 59(11):94--102, 2016.

\bibitem{LangfordZ07}
J.~Langford and T.~Zhang.
\newblock The epoch-greedy algorithm for multi-armed bandits with side
  information.
\newblock In {\em NIPS}, 2007.

\bibitem{li2012unbiased}
L.~Li, W.~Chu, J.~Langford, T.~Moon, and X.~Wang.
\newblock An unbiased offline evaluation of contextual bandit algorithms with
  generalized linear models.
\newblock {\em JMLR}, 26:19--36, 2012.

\bibitem{li2010contextual}
L.~Li, W.~Chu, J.~Langford, and R.~E. Schapire.
\newblock A contextual-bandit approach to personalized news article
  recommendation.
\newblock In {\em WWW}, pages 661--670. ACM, 2010.

\bibitem{pandey2007bandits}
S.~Pandey, D.~Agarwal, D.~Chakrabarti, and V.~Josifovski.
\newblock Bandits for taxonomies: A model-based approach.
\newblock In {\em SDM}, pages 216--227. SIAM, 2007.

\bibitem{pandey2007multi}
S.~Pandey, D.~Chakrabarti, and D.~Agarwal.
\newblock Multi-armed bandit problems with dependent arms.
\newblock In {\em ICML}, pages 721--728. ACM, 2007.

\bibitem{riquelme2018deep}
C.~Riquelme, G.~Tucker, and J.~Snoek.
\newblock Deep bayesian bandits showdown: An empirical comparison of bayesian
  deep networks for thompson sampling.
\newblock {\em arXiv preprint arXiv:1802.09127}, 2018.

\bibitem{schein2002methods}
A.~I. Schein, A.~Popescul, L.~H. Ungar, and D.~M. Pennock.
\newblock Methods and metrics for cold-start recommendations.
\newblock In {\em Proceedings of the 25th annual international ACM SIGIR
  conference on Research and development in information retrieval}, pages
  253--260. ACM, 2002.

\bibitem{smith2013sequential}
A.~Smith, A.~Doucet, N.~de~Freitas, and N.~Gordon.
\newblock {\em Sequential Monte Carlo methods in practice}.
\newblock Springer Science \& Business Media, 2013.

\bibitem{tang2015personalized}
L.~Tang, Y.~Jiang, L.~Li, C.~Zeng, and T.~Li.
\newblock Personalized recommendation via parameter-free contextual bandits.
\newblock In {\em SIGIR}, pages 323--332. ACM, 2015.

\bibitem{tokic2010adaptive}
M.~Tokic.
\newblock Adaptive $\varepsilon$-greedy exploration in reinforcement learning
  based on value differences.
\newblock In {\em KI 2010: Advances in Artificial Intelligence}, pages
  203--210. Springer, 2010.

\bibitem{wang2017factorization}
H.~Wang, Q.~Wu, and H.~Wang.
\newblock Factorization bandits for interactive recommendation.
\newblock In {\em AAAI}, pages 2695--2702, 2017.

\bibitem{wang2017online}
Q.~Wang, C.~Zeng, W.~Zhou, T.~Li, L.~Shwartz, and G.~Y. Grabarnik.
\newblock Online interactive collaborative filtering using multi-armed bandit
  with dependent arms.
\newblock {\em preprint arXiv:1708.03058}, 2017.

\bibitem{wang2017constructing}
Q.~Wang, W.~Zhou, C.~Zeng, T.~Li, L.~Shwartz, and G.~Y. Grabarnik.
\newblock Constructing the knowledge base for cognitive it service management.
\newblock In {\em SCC}, pages 410--417. IEEE, 2017.

\bibitem{wang2017interactive}
X.~Wang, S.~C. Hoi, C.~Liu, and M.~Ester.
\newblock Interactive social recommendation.
\newblock In {\em CIKM}, pages 357--366. ACM, 2017.

\bibitem{wu2016contextual}
Q.~Wu, H.~Wang, Q.~Gu, and H.~Wang.
\newblock Contextual bandits in a collaborative environment.
\newblock In {\em Proceedings of the 39th International ACM SIGIR conference on
  Research and Development in Information Retrieval}, pages 529--538. ACM,
  2016.

\bibitem{yue2012hierarchical}
Y.~Yue, S.~A. Hong, and C.~Guestrin.
\newblock Hierarchical exploration for accelerating contextual bandits.
\newblock {\em arXiv preprint arXiv:1206.6454}, 2012.

\bibitem{zeng2014hierarchical}
C.~Zeng, T.~Li, L.~Shwartz, and G.~Y. Grabarnik.
\newblock Hierarchical multi-label classification over ticket data using
  contextual loss.
\newblock In {\em 2014 IEEE NOMS}, pages 1--8. IEEE, 2014.

\bibitem{tang2017integrated}
C.~Zeng, L.~Tang, W.~Zhou, T.~Li, L.~Shwartz, G.~Grabarnik, et~al.
\newblock An integrated framework for mining temporal logs from fluctuating
  events.
\newblock {\em IEEE Transactions on Services Computing}, 2017.

\bibitem{zeng2016online}
C.~Zeng, Q.~Wang, S.~Mokhtari, and T.~Li.
\newblock Online context-aware recommendation with time varying multi-armed
  bandit.
\newblock In {\em SIGKDD}, pages 2025--2034. ACM, 2016.

\bibitem{zeng2016onlinebigdata}
C.~Zeng, Q.~Wang, W.~Wang, T.~Li, and L.~Shwartz.
\newblock Online inference for time-varying temporal dependency discovery from
  time series.
\newblock In {\em Big Data (Big Data), 2016 IEEE International Conference on},
  pages 1281--1290. IEEE, 2016.

\bibitem{zeng2017knowledge}
C.~Zeng, W.~Zhou, T.~Li, L.~Shwartz, and G.~Y. Grabarnik.
\newblock Knowledge guided hierarchical multi-label classification over ticket
  data.
\newblock {\em IEEE TNSM}, 2017.

\bibitem{zhao2013interactive}
X.~Zhao, W.~Zhang, and J.~Wang.
\newblock Interactive collaborative filtering.
\newblock In {\em CIKM}, pages 1411--1420. ACM, 2013.

\bibitem{zheng2018drn}
G.~Zheng, F.~Zhang, Z.~Zheng, Y.~Xiang, N.~J. Yuan, X.~Xie, and Z.~Li.
\newblock Drn: A deep reinforcement learning framework for news recommendation.
\newblock 2018.

\bibitem{zhou2016latent}
L.~Zhou and E.~Brunskill.
\newblock Latent contextual bandits and their application to personalized
  recommendations for new users.
\newblock {\em arXiv preprint arXiv:1604.06743}, 2016.

\bibitem{zhou2017star}
W.~Zhou, W.~Xue, R.~Baral, Q.~Wang, C.~Zeng, T.~Li, J.~Xu, Z.~Liu, L.~Shwartz,
  and G.~Ya~Grabarnik.
\newblock Star: A system for ticket analysis and resolution.
\newblock In {\em SIGKDD}, pages 2181--2190. ACM, 2017.

\end{thebibliography}
%
\balancecolumns
\end{document}